\documentclass[final,5p,times,twocolumn]{elsarticle}

\usepackage{amsmath,amssymb,amsfonts}
\usepackage{algorithmic}
\usepackage{algorithm}
\usepackage{array}
\usepackage{textcomp}
\usepackage{stfloats}
\usepackage{url}
\usepackage{verbatim}
\usepackage{graphicx}
\usepackage{multirow}
\usepackage{subfigure}
\usepackage{threeparttable}
\usepackage{float}
\usepackage{booktabs}
\usepackage{notoccite}
\usepackage{longtable}
\usepackage{makecell}
\usepackage{lineno}
\usepackage{color}
\usepackage{enumitem}
\usepackage{caption}

\journal{Neural Networks}

\begin{document}

\begin{frontmatter}

\title{A Temporal-Spectral Fusion Transformer with Subject-Specific Adapter for Enhancing RSVP-BCI Decoding}

\author[label1,label2]{Xujin Li \fnmark[1]}
\author[label1]{Wei Wei \fnmark[1]}
\author[label1]{Shuang Qiu \corref{cor1}}
\author[label1,label2]{Huiguang He \corref{cor1}} 
\cortext[cor1]{Corresponding authors: Huiguang He, Shuang Qiu}
\ead{huiguang.he@ia.ac.cn}
\fntext[fn0]{The first two authors contributed equally to this work.}

\affiliation[label1]{organization={Laboratory of Brain Atlas and Brain-Inspired Intelligence, State Key Laboratory of Multi-modal Artificial Intelligence Systems, Institute of Automation, Chinese Academy of Sciences},
            city={Beijing},
            postcode={100190},  
            country={China}}

\affiliation[label2]{organization={School of Future Technology, University of Chinese Academy of Sciences (UCAS)},
            city={Beijing},
            postcode={100049}, 
            country={China}}

\begin{abstract}
    The Rapid Serial Visual Presentation (RSVP)-based Brain-Computer Interface (BCI) is an efficient technology for target retrieval using electroencephalography (EEG) signals. The performance improvement of traditional decoding methods relies on a substantial amount of training data from new test subjects, which increases preparation time for BCI systems. Several studies introduce data from existing subjects to reduce the dependence of performance improvement on data from new subjects, but their optimization strategy based on adversarial learning with extensive data increases training time during the preparation procedure. Moreover, most previous methods only focus on the single-view information of EEG signals, but ignore the information from other views which may further improve performance. To enhance decoding performance while reducing preparation time, we propose a \textbf{T}emporal-\textbf{S}pectral fusion trans\textbf{former} with \textbf{S}ubject-specific \textbf{A}dapter (TSformer-SA). Specifically, a cross-view interaction module is proposed to facilitate information transfer and extract common representations across two-view features extracted from EEG temporal signals and spectrogram images. Then, an attention-based fusion module fuses the features of two views to obtain comprehensive discriminative features for classification. Furthermore, a multi-view consistency loss is proposed to maximize the feature similarity between two views of the same EEG signal. Finally, we propose a subject-specific adapter to rapidly transfer the knowledge of the model trained on data from existing subjects to decode data from new subjects. Experimental results show that TSformer-SA significantly outperforms comparison methods and achieves outstanding performance with limited training data from new subjects. This facilitates efficient decoding and rapid deployment of BCI systems in practical use.
\end{abstract}


\end{frontmatter}

\section{Introduction}
    The Brain-Computer Interface (BCI) systems facilitate the translation of neural signals into computer commands, establishing a direct pathway for communication between the human brain and external devices \cite{wolpaw2002brain}. In BCI systems, electroencephalography (EEG) is generally utilized owing to its high temporal resolution, cost-effectiveness, portability, and noninvasive applicability \cite{nicolas2012brain,ang2013brain}. Rapid Serial Visual Presentation (RSVP)-based BCI systems have attracted significant attention due to their potential for enhancing human-computer collaboration \cite{marathe2015effect,lees2018review} and have been applied in various fields, including target image retrieval \cite{alpert2013spatiotemporal,marathe2015improved}, spellers \cite{acqualagna2013gaze}, face recognition \cite{chen2024eeg}, anomaly detection \cite{barngrover2015brain}, and anti-deception \cite{wu2018anti}.
    
\begin{figure}[!htbp]
		\centering
		\includegraphics[width=\linewidth]{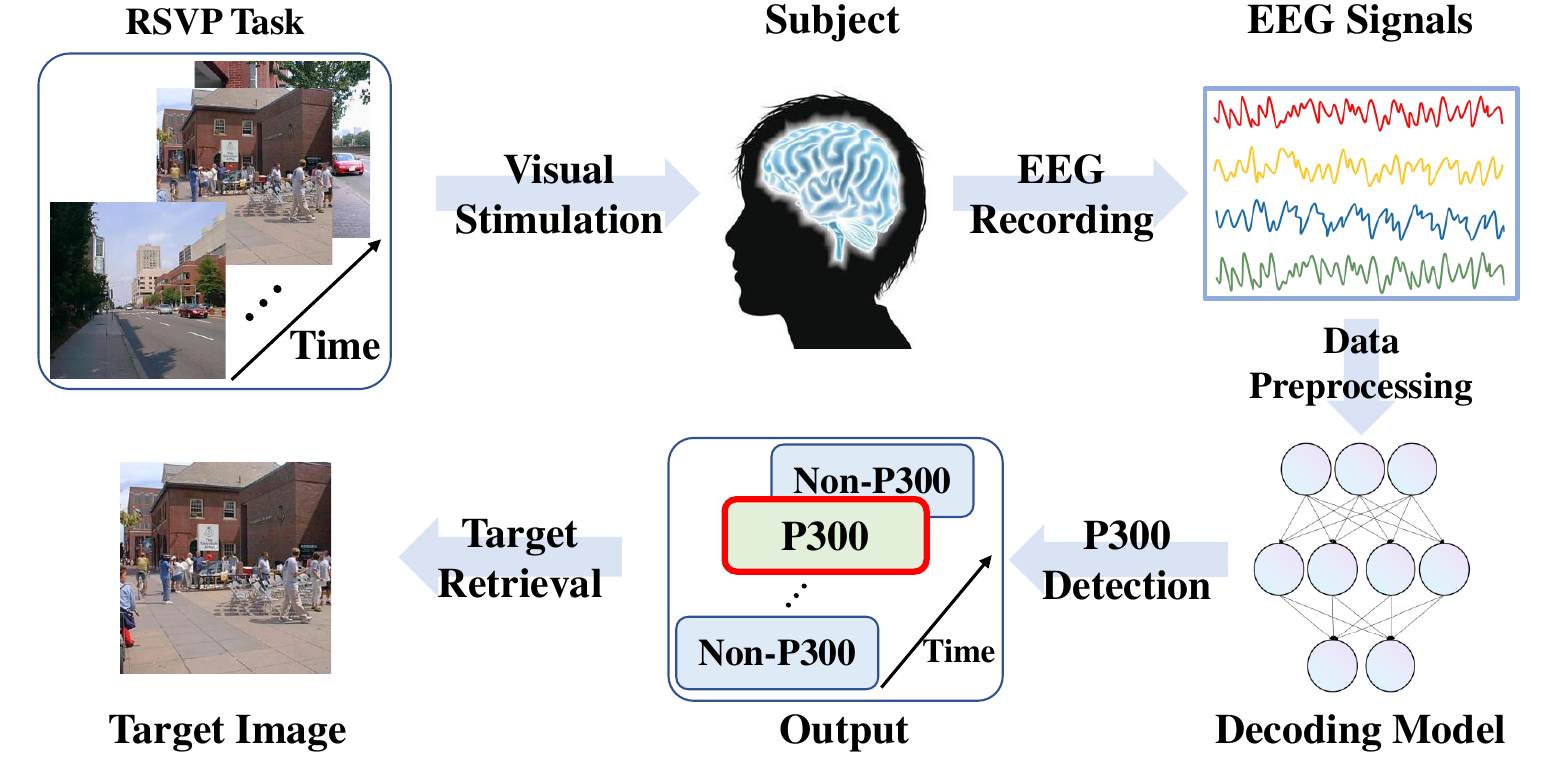}
		\caption{The flowchart of the RSVP-BCI system for target retrieval. The image sequence is presented to the subject at a high rate and the EEG signals of the subject are recorded simultaneously. Subsequently, the RSVP decoding model detects EEG signals containing P300 components, and their corresponding stimulus images are the target images.}
        \label{RSVP system}
	\end{figure}

    In the RSVP-BCI system, an image sequence including various classes is presented to a subject at a high rate \cite{cecotti2014single}. The subject needs to identify the specific target images from other distractors (i.e. nontarget images), where the target images are sparsely distributed in a random order. As a result, these rare target images evoke event-related potentials (ERPs) containing the P300 component which is a prominent positive peak typically observed around 300 ms after infrequent stimuli \cite{squires1976effect,polich2007updating}. The retrieval of the target images can be realized through the detection of the P300 component within EEG signals (see Fig. \ref{RSVP system}).

    P300 components in the same task are relatively consistent in terms of both amplitude and latency, but it is still challenging to achieve single-trial RSVP decoding. Many algorithms have been developed based on conventional machine learning methods, which are applicable to RSVP decoding. In 2006, Gerson et al. proposed a linear discriminant method called Hierarchical Discriminant Component Analysis (HDCA) which operates on channel and time dimensions of EEG signals \cite{HDCA}. Barachant et al. introduced Riemann geometry to measure the distance between EEG raw signals and proposed the Minimum Distance to Riemann Mean (MDRM) \cite{MDRM}. These methods achieve RSVP decoding by leveraging the temporal domain characteristics of EEG temporal signals. Since ERPs can induce phase-synchronized oscillations in EEG signals, resulting in event-related changes in both amplitude and phase at specific frequencies \cite{makeig2002dynamic,makeig2004mining}, researchers also utilize the time-frequency characteristics of EEG signals for P300 detection \cite{robbins2020sensitive}. Several methods have employed Continuous Wavelet Transform (CWT) to convert EEG temporal signals into spectrogram images for extracting spectral features and subsequently classified them using either Linear Discriminant Analysis (LDA) or Support Vector Machine (SVM) \cite{bostanov2004bci,roach2008event}.
    
    With the remarkable progress achieved by deep learning in numerous visual tasks \cite{krizhevsky2012imagenet,lecun2015deep}, deep learning has also been demonstrated to be useful in decoding EEG signals of RSVP tasks. Many deep learning methods, including MCNN \cite{MCNN}, EEGNet \cite{EEGNet}, One Convolution Layer Neural Network (OCLNN) \cite{OCLNN}, Phase-locked Network (PLNet) \cite{PLNet}, and Phase Preservation Neural Network (PPNN) \cite{PPNN}, have improved the RSVP decoding performance by distinguishing temporal domain characteristics of EEG signals. Moreover, several methods that utilize EEG time-frequency information have been developed. For instance, Tajmirriahi et al. proposed an interpretable Convolutional Neural Network (CNN) for P300 detection by leveraging time-frequency characteristics \cite{tajmirriahi2022interpretable}. Havaei et al. also incorporated the time-frequency localization properties of the Gabor transform and CNN to distinguish P300 signals \cite{havaei2023efficient}. Because EEG signals can vary significantly across subjects and over time arising from factors such as mental states, the spatial geometry of the head, and the diversity of brain functions \cite{morioka2015learning}, the aforementioned methods require the collection of labeled data from each new subject on the same day to train individual models, which can be called the subject-dependent decoding. 

    Enhancing the performance of deep learning models necessitates an extensive training dataset \cite{li2022tff}. However, collecting a large amount of training data from the new subject before using the BCI system requires a time-consuming preparation procedure, which can delay the application of the RSVP-BCI system and lead to subject fatigue. This limits the potential for performance improvement in RSVP-BCI systems. To further improve RSVP decoding performance, researchers have attempted to use labeled data collected from existing subjects and a small amount of data from the new subject to jointly train the model to obtain better RSVP decoding performance on the new subject \cite{wei2020reducing}. 
    Wei et al. introduced a conditional adversarial network that aligns the feature distribution of each existing subject with that of the new subject, enabling the feature extractor trained on data from existing subjects to extract task-related features from data of the new subject \cite{wei2020reducing}. Similarly, Fan et al. utilized the adversarial network to mitigate the distribution discrepancy between existing subjects and the new subject \cite{fan2022dc}. These methods utilize the limited training data from new subjects to achieve better performance than comparison methods, thus reducing the requirement for training data from new subjects.

    However, previous RSVP decoding methods still have two limitations for application. Firstly, most previous methods only focus on the single-view information of EEG signals while ignoring the discriminative information from other views in the process of designing feature extractors. This hinders the extraction of comprehensive EEG representations, which may limit the further improvement of decoding performance. Secondly, models that introduce existing subjects' data are trained using substantial quantities of data from both existing and new subjects through adversarial learning, which significantly increases the training time and leads to unstable and slow convergence during model training. Consequently, these methods fail to effectively reduce the preparation time required before new subjects can utilize the RSVP-BCI system. 

    \begin{figure}[!htbp]
		\centering
		\includegraphics[width=0.9\linewidth]{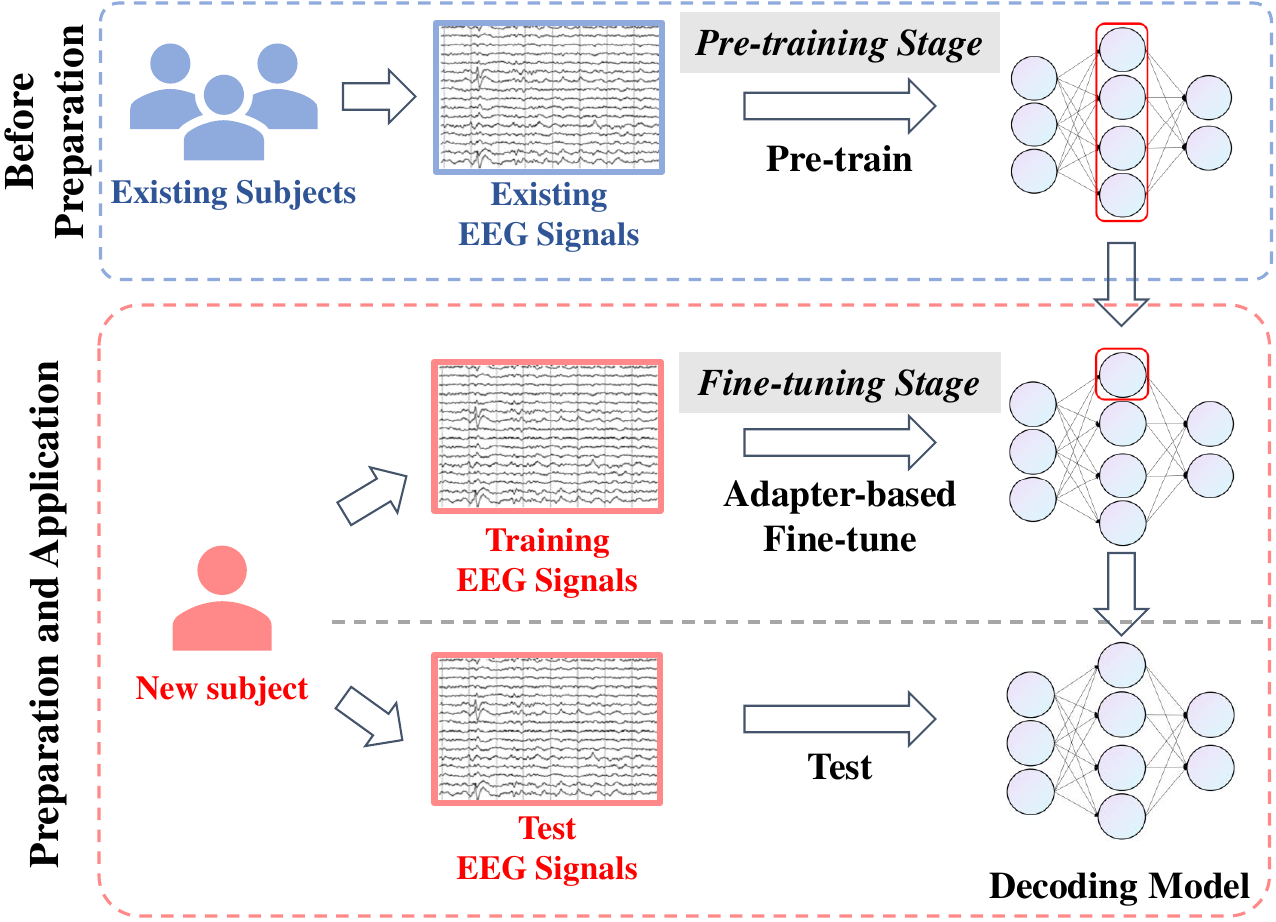}
		\caption{The framework of our proposed two-stage training strategy. The model is initially pre-trained on the EEG signals from existing subjects before the preparation procedure of the BCI system. During the preparation procedure, the training EEG signals from the new subject are used to fine-tune the adapter of the model. Then the model is utilized to decode EEG signals from the new subject.}
        \label{two stage training}
	\end{figure}
   
    In this paper, to enhance RSVP decoding performance while reducing preparation time for RSVP-BCI systems, we propose a Temporal-Spectral fusion transformer with Subject-specific  Adapter (TSformer-SA) which is optimized through a two-stage training strategy. Inspired by studies that incorporate multi-view information in EEG decoding and obtain more discriminative high-level representations \cite{tian2019deep,sun2022dual,liu20233d}, we integrate the temporal and spectral information of EEG signals within a multi-view learning network based on Transformer, which is compatible with diverse modalities and can realize cross-view interaction \cite{xu2023multimodal}. The EEG temporal signals representing the temporal view and spectrogram images obtained through CWT representing the spectral view are utilized for input. Firstly, we propose a feature extractor to tokenize the input and extract global features within each view. Secondly, the cross-view interaction module and the multi-view consistency loss are designed to reduce the disparity between the two-view features from the same EEG signal for extracting common task-related representations. Thirdly, we propose an attention-based fusion module to effectively fuse the features extracted from two views to obtain discriminative features containing both temporal and spectral characteristics for classification. Moreover, the adapter-based fine-tuning can transfer learned general patterns of pre-trained models to downstream tasks and consequently enhance decoding performance by updating a few parameters \cite{houlsby2019parameter}. Thus, instead of training the model from scratch with substantial quantities of existing data, we propose the subject-specific adapter and design a two-stage training approach to leverage the data from existing subjects while reducing the preparation time. In the two-stage training approach, we employ the data from existing subjects to pre-train the TSformer in advance and only fine-tune the subject-specific adapter using data from the new subject in the preparation procedure (see Fig. \ref{two stage training}). The subject-specific adapter is utilized to adapt the model pre-trained on the data from existing subjects to the data characteristics of new subjects and enhance decoding performance with limited training data from new subjects. The main contributions of this work are summarized as follows: 
    \begin{itemize}
        \item[1)] We propose a novel Temporal-Spectral fusion transformer with Subject-specific Adapter (TSformer-SA) optimized through a two-stage training strategy to enhance RSVP decoding performance and reduce the preparation time of BCI systems. The codes have been released at: https://github.com/lixujin99/TSformer-SA.
        
        \item[2)] We propose the cross-view interaction module based on cross-attention and token fusion mechanisms to extract common representations across temporal and spectral views of EEG signals. Furthermore, we also propose the multi-view consistency loss to further narrow the gap between two-view features.
     
        \item[3)] An efficient subject-specific adapter is proposed to realize rapid adaptation of the model to the data characteristics of new subjects, which can enhance the RSVP decoding performance and reduce the model training time during the preparation procedure. 
        
        \item[4)] Extensive experiments are conducted on our self-collection open-source dataset to verify the performance of TSformer-SA. The experimental results demonstrate that our model achieves excellent decoding performance while reducing the requirement for new subjects' training data and the training time in the preparation procedure.
    \end{itemize}
 
    \section{Related Work}
    This section provides a brief review of previous studies related to our proposed method, including RSVP decoding methods, Transformer-based EEG decoding methods, and multi-view learning EEG decoding methods. The RSVP decoding methods are further categorized into conventional machine learning methods and CNN-based methods.

    \subsection{RSVP Decoding Methods}
    \subsubsection{Conventional Machine Learning Methods}
    In 2006, Gerson, A.D. et al. introduced HDCA, which employed linear discriminant analysis to compute spatial weighting vectors for various time intervals and project single-trial EEG data. This method has been widely adopted in RSVP decoding for its simplicity and efficiency \cite{HDCA}. In 2014, Barachan et al. proposed the MDRM method \cite{MDRM}, which was rooted in Riemannian geometry and classified covariance matrices derived from original EEG signals based on their minimum distance to the mean. In 2016, Waytowich et al. introduced Spectral Transfer Learning using Information Geometry (STIG) for subject-independent RSVP decoding, which was also based on Riemannian geometry. The STIG employed spectral-meta learning to ensemble information geometry classifiers trained on data from other training subjects for classification \cite{waytowich2016spectral}.

    \subsubsection{CNN-Based Methods}
    In recent years, deep learning methods have gained increasing popularity for classifying EEG data and demonstrated superior decoding performance. In 2011, Cecotti et al. first introduced the CNN for P300 detection. The network incorporated spatial and temporal convolution layers for feature extraction and included multiple model variants, which proves the effectiveness of CNN for RSVP decoding \cite{cecotti2010convolutional}. Subsequently, Manor et al. (2015) applied CNN to the RSVP target image retrieval task and proposed a larger network combining a novel spatiotemporal regularization method specifically designed for EEG signals \cite{MCNN}. In 2018, Lawhern Vernon J et al. introduced EEGNet, a compact CNN incorporating depthwise separable convolution \cite{chollet2017xception} which can simplify the structure and reduce the number of parameters \cite{EEGNet}. In 2020, Lee et al. introduced a model based on EEGNet for the subject-independent P300 BCI speller, which demonstrated comparable performance to subject-dependent methods \cite{LeeCNN}. In 2021, Zang et al. leveraged the phase-locked characteristic of ERPs and proposed PLNet which separately processes spatial convolution in different time periods to extract spatiotemporal features for RSVP decoding. \cite{PLNet}. Meanwhile, Li et al. proposed a phase preservation neural network (PPNN) to learn phase information of EEG signals in all frequency bands and improve the classification performance in the RSVP task \cite{PPNN}.
	
    \subsection{Transformer-Based EEG Decoding Methods} 
    In comparison to convolutions which utilize identical filter weights for all input patches, Transformer is an attention-based structure that can flexibly adjust the receptive field and effectively capture global interactions via self-attention mechanism \cite{ramachandran2019stand,naseer2021intriguing}. Over the past two years, researchers have explored the use of Transformers in EEG decoding. For SSVEP classification, Chen et al. (2023) proposed the SSVEPformer \cite{chen2023transformer}, which leveraged complex spectrum features extracted from SSVEP data as input and explored spectral and spatial information. In Motor Imagery (MI) decoding, Song et al. (2023) proposed the EEG-conformer, a compact convolutional Transformer capable of unifying local and global features within a single EEG classification framework \cite{song2022eeg}. In the emotion recognition task, Casal et al. (2022) employed a temporal convolution network to extract original features as the input of the Transformer for learning the transition rules among sleep stages \cite{TCN-T}. This model can be applied to both heart rate and EEG decoding tasks \cite{li2022tff}. In 2024, Yan et al. combined a graph convolution network and multi-scale transformer for EEG emotion recognition, where the graph-based networks primarily focused on local EEG channel relationships and the multi-scale transformer learned spatial features more elaborately to enhance feature representation ability \cite{yan2024bridge}. All of these studies demonstrate the effectiveness of Transformers in EEG decoding. 
   
    \subsection{Multi-View Learning EEG Decoding Methods}
    Multi-view representation learning tackles the challenge of learning effective representations from multi-view data for enhancing the development of prediction models \cite{li2018survey}. In 2021, Li et al. proposed a novel temporal-spectral-based squeeze-and-excitation feature fusion network, which effectively fused temporal features extracted by temporal convolutions and spectral features extracted by multi-level wavelet convolutions for decoding MI-EEG signals \cite{li2021temporal}. The next year, Sun et al. proposed a model based on dynamic graph convolution and adaptive feature fusion \cite{sun2022dual}. This model computed both differential entropy and power spectral density features from each EEG segment as two distinct views, and a dual-branch graph dynamic convolution (DBGC) network was designed to capture deeper temporal and spectral features. In 2023, Luo et al. proposed a multi-view fusion model based on Transformer for EEG-based visual recognition \cite{luo2023dual}, which fused features from the spatio-temporal and spectral-temporal domains of EEG signals and achieved superior performance compared to state-of-the-art methods. These approaches demonstrate that utilizing multi-view information of EEG signals enables the model to extract more discriminative high-level features, which enhances the classification performance.

    \section{Materials}
    \subsection{RSVP Paradigm}
    In this study, we aim to propose a method to enhance the RSVP decoding performance and evaluate its performance in different practical application scenarios. Therefore, we design three different RSVP experiments to conduct three independent datasets. 
    
    \subsubsection{RSVP Tasks}
    \begin{figure}[!htbp]
        \centering
        \subfigure[]{
			\includegraphics[width=0.95\linewidth]{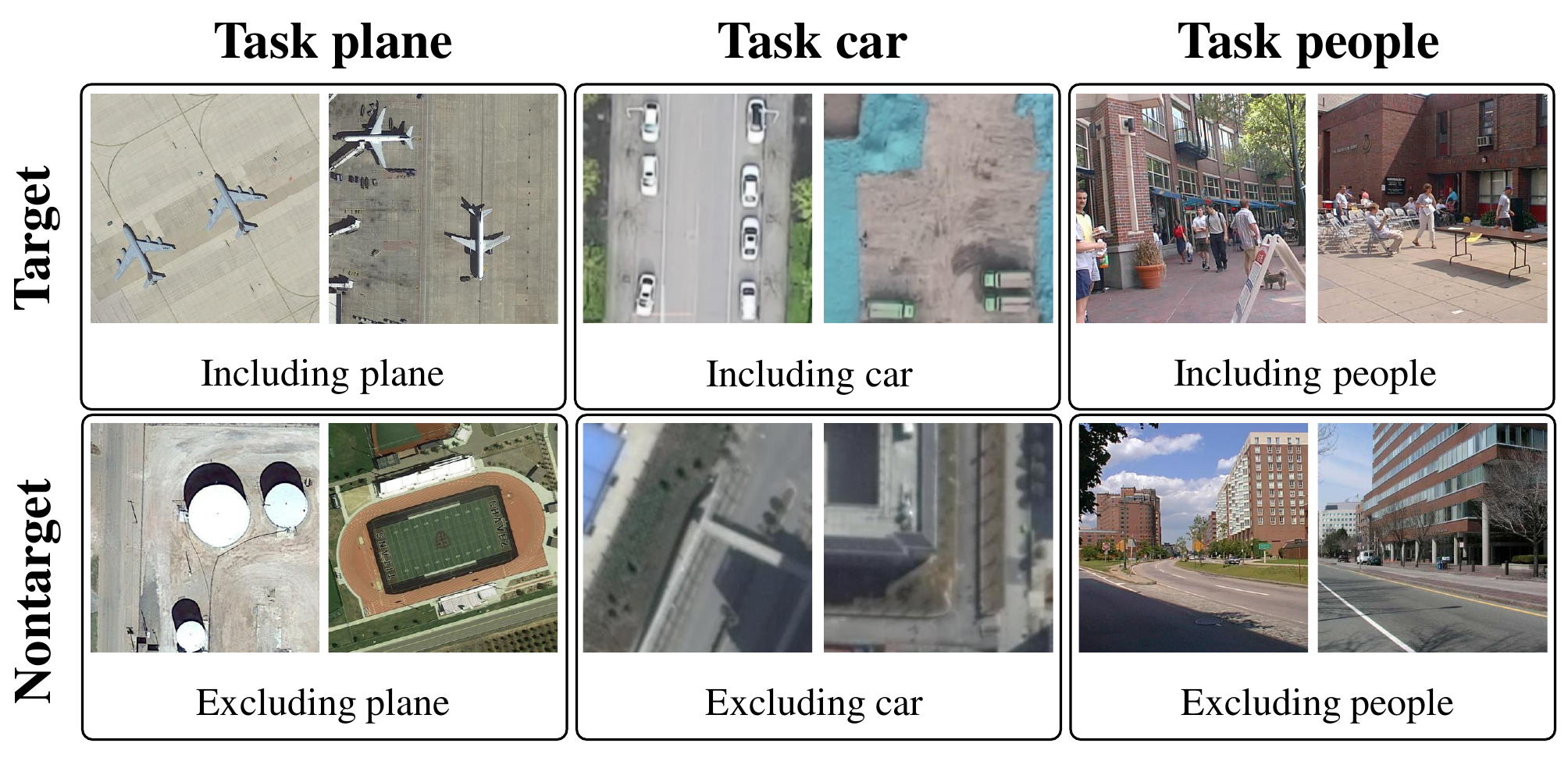}
		}
		\subfigure[]{
			\includegraphics[width=0.98\linewidth]{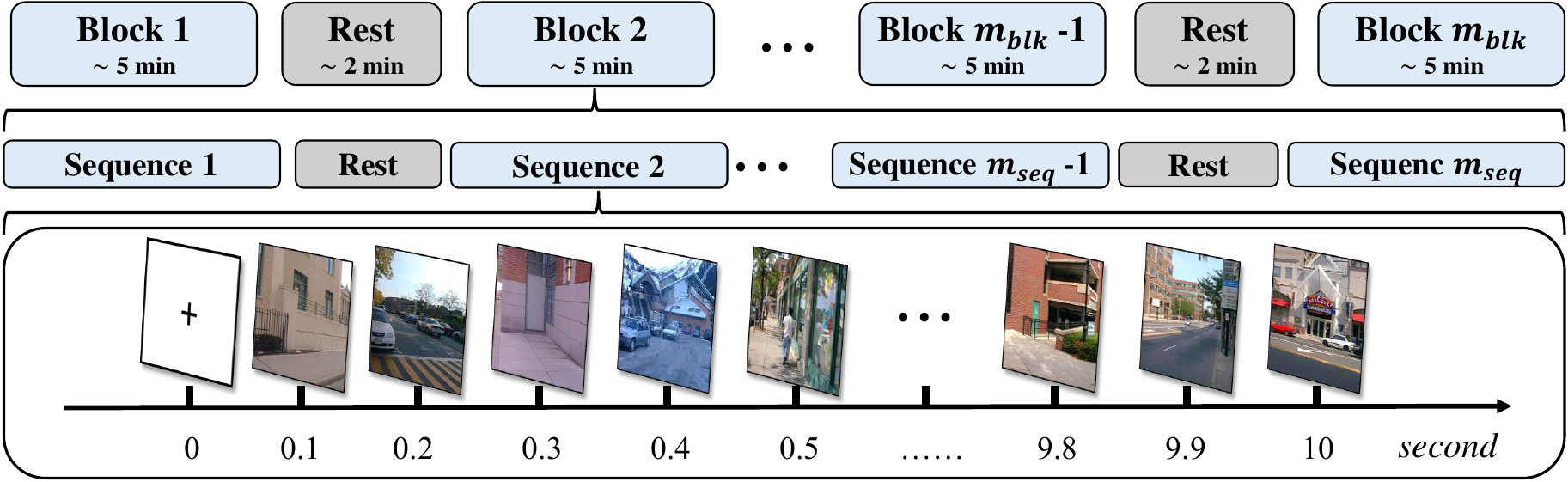}
		}
		\caption{Illustration of the RSVP paradigm. (a) Examples of target and nontarget images in Task plane, Task car, and Task people. The stimulus images in Task plane are sourced from the remote sensing Dior dataset \cite{li2020object}, the stimulus images of the Task car are from our self-collection drone aerial images, and the stimulus images of the Task people are from the scenes and objects database \cite{torralba2009csail}. (b) Experimental settings about the division of blocks and sequences for each subject.}
        \label{RSVP experiment}
    \end{figure}

    At present, retrieving images containing target people from street view images is the most commonly used RSVP task in previous studies \cite{cecotti2010convolutional,alpert2013spatiotemporal,marathe2015improved,mao2023cross,zhang2020benchmark}, which has been verified the validity of this paradigm \cite{waytowich2016spectral}. Hence, in this study, we select the scenes and objects database \cite{torralba2009csail} to conduct an RSVP task that retrieves target people from street view images to evaluate the model performance. To enhance the diversity of RSVP tasks in our study and focus on practical applications, we also design two RSVP tasks to simulate optical remote sensing observation \cite{miranda2015darpa,matran2016brain} and drone observation \cite{zhang2023uav} respectively. We utilize the remote sensing Dior dataset \cite{li2020object} and our self-collection drone aerial images to simulate observation, respectively. Considering that the majority of our subjects are college students, we choose common objects that are familiar to the subjects and have high practical value as targets, so that the subjects can start the target retrieval task with simple training duration. Hence, in the remote sensing observation, we select the images containing a plane in the Dior dataset as the target, and we select the car as the target of drone observation in the drone aerial images. 
    
    As a result, we have designed three distinct RSVP tasks: (1) Task plane involves identifying planes from remote sensing images, which is designed to simulate remote sensing observation; (2) Task car focuses on identifying cars from images captured by drones, which simulates the drone observation; (3) Task people involves retrieving images containing people from street scene images, which simulates the security monitor in street view. Within each task, the visual stimulus images that contain the task-specified target are designated as target images, while the remaining images are considered nontarget images. Figure \ref{RSVP experiment} (a) illustrates a few examples of the stimulus images. Finally, we recruit participants for each of the three RSVP tasks respectively.

    \subsubsection{Parameters of RSVP Paradigm}
    The images in each RSVP task are randomly presented at a frequency of 10 Hz, with the target images appearing with a probability of 4\%. Participants are seated 1 meter away from the monitor with a resolution of 1920 x 1080. When participants achieve a state of calmness, they are presented with a sequence of images and instructed to identify the target images. As shown in Fig. \ref{RSVP experiment} (b), each subject completes a task consisting of $m_{blk}$ blocks (Task plane: $m_{blk}=10$; Task car: $m_{blk}=5$; Task people: $m_{blk}=10$). Each block contains $m_{seq}$ sequences (Task plane: $m_{seq}=14$; Task car: $m_{seq}=16$; Task people: $m_{seq}=14$), where each sequence consists of 100 images. The interval between sequences is controlled by the participants. Each block takes around 5 minutes to complete, with a break time of approximately 2-3 minutes between blocks.

    \subsection{Subjects}
    In the experiment, we design and implement three RSVP tasks for target image retrieval including Task plane, Task car, and Task people. Each task is an independent dataset. The Task plane involves 20 subjects (15 males and 5 females, aged 23.4 ± 1.1, with 18 being right-handed). The Task car involves 20 subjects (11 males and 9 females, aged 23.75 ± 1.3, all of whom are right-handed). The Task people involves 31 subjects (19 males and 12 females, aged 24.9 ± 2.8, with 28 being right-handed). 
    
    There is no overlap in the subject participation across the three tasks. All participants possess normal or corrected-to-normal vision. Before the experiment, the Institutional Review Board of the Institute of Automation, Chinese Academy of Sciences, provides ethical approval. Additionally, all subjects are voluntary and provide written informed consent.
	
    \subsection{Data Acquisition and Preprocessing}
    The data acquisition and preprocessing procedures are consistent for all three tasks. EEG data are recorded using the SynAmp2 system (Neuroscan, Australia) with 64-channel Ag/AgCl electrodes placed according to the international 10/20 system, with a sampling rate of 1000 Hz. All the electrodes with an impedance of 10 k$\Omega$ or lower are referenced to the vertex and grounded on the AFz channel. 

    During the preprocessing stage, the EEG data are downsampled to 250 Hz. Subsequently, a linear phase 3-order Butterworth filter with a bandpass ranging from 0.5 to 15 Hz is applied to the signal to eliminate slow drift and high-frequency noise and prevent delay distortions. Next, the preprocessed data from each block are segmented into EEG trials. Each trial consists of 1-second EEG data starting from the stimulation onset to 1000 milliseconds (0 s to 1 s) after the stimulation onset. For each trial, data are normalized to zero mean and variance one. The subsequent analysis and classification of EEG signals rely on these segmented EEG trials (samples). Based on our experimental paradigm, each subject has a total of $m_{blk}\times m_{seq}\times 100$ EEG samples.

    \begin{figure*}[htbp]
		\centering
		\includegraphics[width=0.92\linewidth]{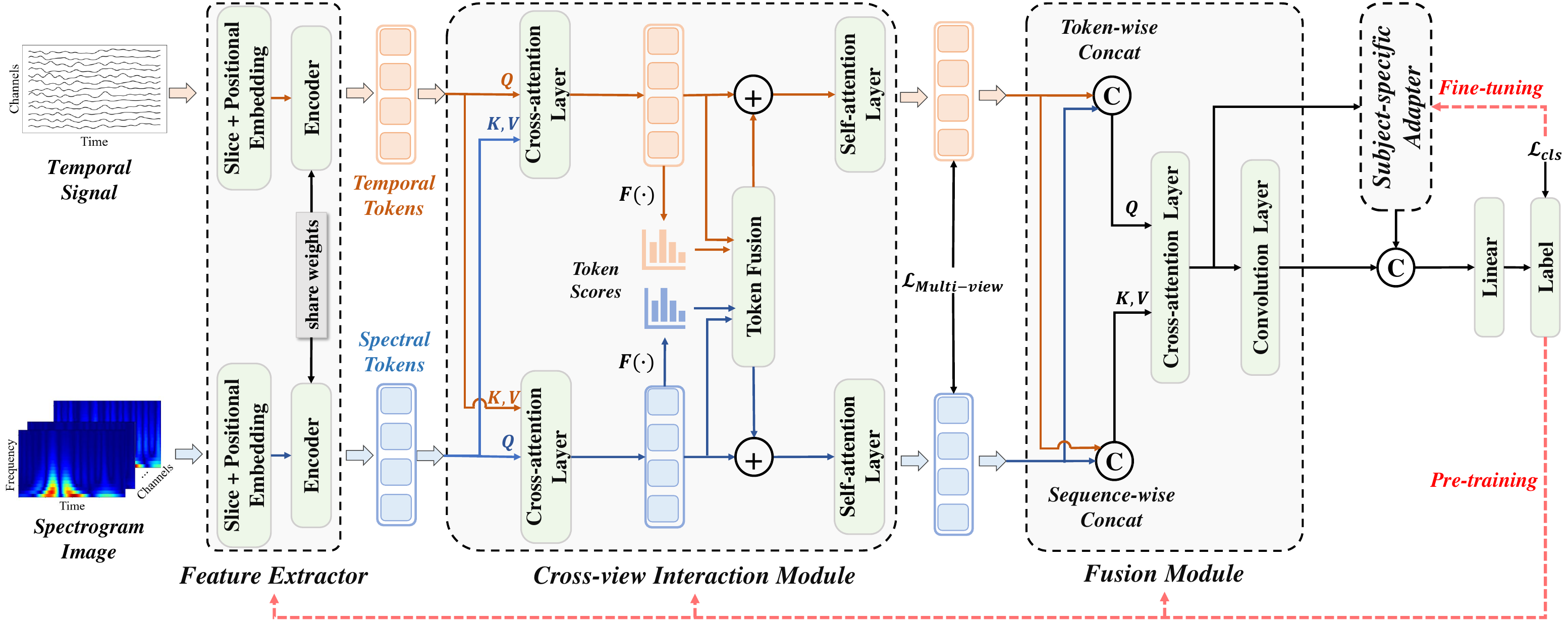}
        
		\caption{The structure of our proposed TSformer-SA. The $F(\cdot)$ represents the token score function, while $\mathcal{L}_{cls}$ and $\mathcal{L}_{multi\mbox{-}view}$ respectively denote the cross-entropy loss and the multi-view consistency loss. The inputs consist of EEG temporal signals representing the temporal view and the spectrogram images representing the spectral view. The feature extractor tokenizes the inputs and extracts the view-specific features. Subsequently, the cross-view interaction module extracts the common features from both views and the fusion module fuses the two-view features for classification. The above three modules are trained during the pre-training stage and only the subject-specific adapter is trained in the fine-tuning stage.}
        \label{model framework}
    \end{figure*}
 
    \section{Method}
    In this section, we introduce the Temporal-Spectral fusion transformer with Subject-specific Adapter (TSformer-SA) for RSVP decoding. As shown in Fig. \ref{model framework}, TSformer-SA is a symmetrical dual-stream Transformer comprising a feature extractor, a cross-view interaction module, a fusion module, and a subject-specific adapter. The inputs of the model are EEG temporal signals as the temporal view and the spectrogram image of each channel computed using CWT as the spectral view. First, the feature extractor employs a slice embedding layer to tokenize the two-view input and an encoder layer to extract view-specific features. Subsequently, the cross-view interaction module employs cross-attention and token fusion mechanisms to realize information transfer between the two views. Simultaneously, we introduce a multi-view consistency loss for the common feature extraction between two views by maximizing the similarity of different view features from the same sample. Then, an attention-based fusion module achieves the feature fusion between two views to obtain comprehensive features for classification. 
    
    The optimization strategy of TSformer-SA consists of the pre-training stage and the adapter-based fine-tuning stage. In the pre-training stage, multi-view consistency loss and cross-entropy loss are used to optimize TSformer on data from existing subjects. Then the proposed subject-specific adapter is fine-tuned on new subjects' data to rapidly transfer the learned general patterns of the pre-trained model to the data distribution of new subjects. To facilitate reading, the frequently used symbols and their definitions are listed Table \ref{summary}.
    
    \setlength{\tabcolsep}{1.1mm}{
	\begin{table}[htbp]
            \footnotesize
            \centering
            \renewcommand\arraystretch{1.3}
            \caption{The frequently used symbols and their definitions.}
            \label{summary}
            \begin{tabular}{clp{5.7cm}}
                \toprule[1.2pt]
                \textbf{Symbol} &\textbf{Shape} & \textbf{Definition} \\
                \hline
                $\boldsymbol{S}_{tem}$&$\mathbb{R}^{1\times C\times T}$ & EEG temporal signals, with $C$ EEG channels and $T$ time points. \\
                \midrule
                $\boldsymbol{S}_{spe}$&$\mathbb{R}^{S\times C\times T}$ & EEG spectrogram images, with $S$ scale. \\
                \midrule
                $\boldsymbol{X}_{tem}, \boldsymbol{X}_{spe}$&$\mathbb{R}^{n\times d}$ & Temporal and spectral token sequences, with $n$ tokens.    \\
                \midrule
                $\boldsymbol{x}_{i}^{tem}$, $\boldsymbol{x}_{i}^{spe} $&$\mathbb{R}^{d}$ & The $i$-th temporal token and spectral token in $\boldsymbol{X}_{tem}$ and $\boldsymbol{X}_{spe}$ respectively, with $d$ dimension.  \\
                \midrule
                $\boldsymbol{z}^{tem}_{i},\boldsymbol{z}^{spe}_{i}$& $\mathbb{R}^{3d}$ & The feature vectors of the temporal view and spectral view from the $i$-th sample used in the multi-view consistency loss. \\
                \midrule
                $\boldsymbol{W}_{Q}, \boldsymbol{W}_{K}, \boldsymbol{W}_{V}$& $\mathbb{R}^{d\times d}$& Learnable parameter matrix in attention mechanisms. \\
                \midrule
                $\boldsymbol{W}_{score}$& $\mathbb{R}^{d\times d}$& Learnable parameter matrix in token score function.\\
                \midrule
                $F(\cdot)$& \textbf{\textemdash} & Token score function.  \\
                \midrule
                $N$& $\mathbb{Z}$& The number of EEG samples.\\
                \bottomrule[1.2pt] 
            \end{tabular}
	\end{table}}

    \subsection{Feature Extractor}
    The feature extractor includes a slice embedding layer and an encoder layer. Since the difference between the P300 signals induced by target images and the harmonic signals induced by non-target images in the RSVP task is concentrated along the time dimension, the preprocessed EEG temporal signals and spectrogram images are initially divided into non-overlapping slices along the time dimension. To notice more valuable time slice information, we adopt the self-attention mechanism within each view to perceive the global time dependencies of EEG temporal signals and spectrogram images.
    
    The inputs of the slice embedding layer are EEG temporal signals ($\boldsymbol{S}_{tem}\in \mathbb{R}^{1\times C\times T}$) and spectrogram images of each channel ($\boldsymbol{S}_{spe}\in \mathbb{R}^{S\times C\times T}$), where $C$ denotes the number of channels, $T$ denotes the number of sampling points, and $S$ denotes the size of scale in CWT. We slice and reshape the input signals $\boldsymbol{S}_{tem}$ and $\boldsymbol{S}_{spe}$ into the sequence of flattened slices $\left[\boldsymbol{s}_{1}^{tem},\boldsymbol{s}_{2}^{tem},\cdots,\boldsymbol{s}_{n}^{tem}\right]$ and $\left[\boldsymbol{s}_{1}^{spe},\boldsymbol{s}_{2}^{spe},\cdots,\boldsymbol{s}_{n}^{spe}\right]$ along the time dimension, respectively. Each $\boldsymbol{s}_{i}^{tem}\in \mathbb{R}^{(C\times t)}$ and $\boldsymbol{s}_{i}^{spe}\in \mathbb{R}^{(S\times C\times t)}$ respectively represents brain temporal and spectral activities over a period of time, where $t$ is the slice length and $n = [T/t]$ is the number of slices. Then, the slices are projected to $d$ dimensions, where the output of these temporal and spectral slice embeddings are referred to as temporal tokens $\boldsymbol{X}_{tem} \in \mathbb{R}^{n\times d}$ and spectral tokens $\boldsymbol{X}_{spe} \in \mathbb{R}^{n\times d}$, respectively. The process can be formulated as follows: 
    \begin{equation}
        \boldsymbol{X}_{tem} =\left[\boldsymbol{s}_{1}^{tem},\boldsymbol{s}_{2}^{tem},\cdots,\boldsymbol{s}_{n}^{tem}\right]^{T}\boldsymbol{W}_{pos}^{(1)} + \boldsymbol{E}_{pos}^{(1)},
    \end{equation}

    \begin{equation}    
        \boldsymbol{X}_{spe} =\left[\boldsymbol{s}_{1}^{spe},\boldsymbol{s}_{2}^{spe},\cdots,\boldsymbol{s}_{n}^{spe}\right]^{T}\boldsymbol{W}_{pos}^{(2)} + \boldsymbol{E}_{pos}^{(2)},
    \end{equation}
    where $\boldsymbol{W}_{pos}^{(i)} \in \mathbb{R}^{(C\times t)\times d}$ ($i=1, 2$) is the parameter matrix and $\boldsymbol{E}_{pos}^{(i)}\in \mathbb{R}^{n\times d}$ ($i=1, 2$) is the learnable positional parameter matrix added to the slice embeddings to retain positional information. The resulting sequence of embedding vectors serves as the input to the encoder layers. 

    The encoder layer comprises two sub-layers: a Multi-Head Self-Attention (MHSA) layer capturing global temporal dependencies within input tokens, and a position-wise Feed-Forward Network (FFN) layer with a hidden layer consisting of $h\times d$ dimensions extracting feature representations, where $h$ denotes the number of heads. All sub-layers utilize a residual connection and layer normalization to enhance the scalability of the model. To reduce model parameters and enable the subsequent view interaction more reasonable and feasible, we share the encoder parameters between the two streams to project the features into the same feature space. The outputs of the feature extractor are temporal tokens ($\boldsymbol{X}_{tem}$) and spectral tokens ($\boldsymbol{X}_{spe}$).
    
    \subsection{Multi-View Interaction Learning}
    To realize view interaction and extract common representations across temporal and spectral views, we employ the cross-view interaction module constrained by multi-view consistency loss for multi-view interaction learning.
    
    \subsubsection{Cross-View Interaction Module}
    The cross-view interaction module consists of a cross-attention layer and a token fusion layer. To capitalize on the complementary information between two views and extract common features, the cross-attention mechanism is first employed to focus on EEG information transfer between temporal and spectral views and adaptively learn inter-view common representations.
    
    The input of the cross-attention layer are token sequences of two views: $\boldsymbol{X}_{tem}=[\boldsymbol{x}^{tem}_{1},\boldsymbol{x}^{tem}_{2},\cdots,\boldsymbol{x}^{tem}_{n}]^{T}\in \mathbb{R}^{n\times d}$ and $\boldsymbol{X}_{spe}=[\boldsymbol{x}^{spe}_{1},\boldsymbol{x}^{spe}_{2},\cdots,\boldsymbol{x}^{spe}_{n}]^{T}\in \mathbb{R}^{n\times d}$, where $\boldsymbol{x}^{tem}_{i}\in \mathbb{R}^{d}$ and $\boldsymbol{x}^{spe}_{i}\in \mathbb{R}^{d}$ denote the $i$-th token in temporal and spectral token sequences, respectively. The linear projection layer maps input token sequences to sequential vectors and then both views mutually extract common features from each other based on the similarity between their tokens. We utilize a multi-head attention mechanism. To concisely represent the model inference process, we illustrate the single-head attention mechanism as an example. The process can be formulated as follows:
    \begin{equation}
            \boldsymbol{X}_{tem}^{CA} = \boldsymbol{X}_{tem} + \sigma(\frac{\boldsymbol{X}_{tem}\boldsymbol{W}_{Q}^{(1)}{\boldsymbol{W}_{K}^{(1)}}^{T}\boldsymbol{X}_{spe}^{T}}{\sqrt{d_{k}}})\boldsymbol{X}_{spe}\boldsymbol{W}_{V}^{(1)},
    \end{equation}
    \begin{equation}
            \boldsymbol{X}_{spe}^{CA} = \boldsymbol{X}_{spe} + \sigma(\frac{\boldsymbol{X}_{spe}\boldsymbol{W}_{Q}^{(2)}{\boldsymbol{W}_{K}^{(2)}}^{T}\boldsymbol{X}_{tem}^{T}}{\sqrt{d_{k}}})\boldsymbol{X}_{tem}\boldsymbol{W}_{V}^{(2)},
    \end{equation}
    where $\boldsymbol{X}_{tem}^{CA} \in \mathbb{R}^{n\times d}$ and $\boldsymbol{X}_{spe}^{CA} \in \mathbb{R}^{n\times d}$ are the temporal and spectral view output of the cross-attention layer respectively. The linear matrices $\boldsymbol{W}_{Q}^{(i)}\in \mathbb{R}^{d\times d}$, $\boldsymbol{W}_{K}^{(i)}\in \mathbb{R}^{d\times d}$, and $\boldsymbol{W}_{V}^{(i)}\in \mathbb{R}^{d\times d}$ ($i=1, 2$) are learnable parameter matrices. The $\sigma(\cdot)$ represents the softmax function.

    As the low signal-to-noise ratio of the single sample EEG signal may lead to redundant tokens embedded from some EEG time periods, it is crucial to dynamically identify and replace these uninformative tokens with features pertinent to decoding tasks. We propose a token score function to measure the importance of each token within each view and then apply token fusion \cite{wang2022multimodal} to selectively prune uninformative tokens and replace each pruned token using projected features at the same time period from alternative views.

    For the updating of temporal tokens $\boldsymbol{X}_{tem}^{CA}$, we assess the importance of token $\boldsymbol{x}^{tem}_{i}$ using a scoring function denoted as $f(\cdot)$:
    \begin{equation}
        f(\boldsymbol{x}^{tem}_{i}) = \sum_{\substack{j=1 \\ j\neq i}}^{n} \sigma \left(\dfrac{\boldsymbol{X}_{tem}^{CA}\boldsymbol{W}_{score}{\boldsymbol{X}_{tem}^{CA}}^{T}}{\sqrt{d}}\right)_{(j,i)},
    \end{equation}
    where $\boldsymbol{W}_{score} \in \mathbb{R}^{d\times d}$ is a learnable parameter matrix. The scoring function represents the accumulation of attention weights from other tokens towards  $\boldsymbol{x}^{tem}_{i}$ in the self-attention matrix of temporal tokens $\boldsymbol{X}_{tem}^{CA}$. Tokens that receive higher attention can be considered to contain more information. On the basis of $f(\cdot)$, we can establish the scoring function of temporal tokens as $F(\cdot)$, which can be formulated as:
    
    \begin{equation}
        F(\boldsymbol{X}_{tem}^{CA}) = \left[f(\boldsymbol{x}^{tem}_{1})\cdot\boldsymbol{1},f(\boldsymbol{x}^{tem}_{2})\cdot\boldsymbol{1},\cdots,f(\boldsymbol{x}^{tem}_{n})\cdot\boldsymbol{1} \right],
    \end{equation}
    where $\boldsymbol{1}\in \mathbb{R}^{d}$ denotes the vector with elements of $1$. Then we use the scoring function $F(\cdot)$ to identify temporal tokens that contain relatively less information. These selected tokens are subsequently averaged with their corresponding spectral tokens that obtain high scores from the same time period, thereby enhancing tokens with restricted information by incorporating information from the other view. This process can be formulated as:
    \begin{equation}
        \boldsymbol{X}_{tem}^{TF} = \dfrac{1}{2} \left[\boldsymbol{X}_{tem}^{CA} + \boldsymbol{X}_{spe}^{CA} \odot \mathbb{I}_{(F(\boldsymbol{X}_{tem}^{CA})<\theta_{tem})} \odot \mathbb{I}_{(F(\boldsymbol{X}_{spe}^{CA})>\theta_{spe})}\right], 
    \end{equation} 
    \begin{equation}
        \boldsymbol{X}_{spe}^{TF} = \dfrac{1}{2} \left[\boldsymbol{X}_{spe}^{CA} + \boldsymbol{X}_{tem}^{CA} \odot \mathbb{I}_{(F(\boldsymbol{X}_{spe}^{CA})<\theta_{spe})} \odot \mathbb{I}_{(F(\boldsymbol{X}_{tem}^{CA})>\theta_{tem})}\right], 
    \end{equation} 
    where the threshold $\theta_{tem}$ and $\theta_{spe}$ are equal to the median of $F(\boldsymbol{X}_{tem}^{CA})$ and $F(\boldsymbol{X}_{spe}^{CA})$, respectively. The $\boldsymbol{X}_{tem}^{TF}, \boldsymbol{X}_{spe}^{TF}\in \mathbb{R}^{n\times d}$ denote the temporal and spectral view output of the token fusion and $\mathbb{I}_{(\cdot)}$ is an indicator asserting the subscript condition, which outputs a mask tensor with element 0 or 1. The operator $\odot$ denotes the element-wise multiplication. Finally, we use a self-attention layer to extract the global features of tokens with added supplementary information.

    \subsubsection{Multi-View Consistency Loss}
    The temporal and spectral features respectively represent the temporal and spectral activities of EEG signals and there are great differences between the two view features. To extract common features from both views to enhance decoding performance, we design a multi-view consistency loss that narrows the distance between the temporal and spectral features from the same sample based on contrastive learning \cite{chen2020simple}.
    
    Firstly, a convolution layer with four convolution kernels, each of size $(16, d/8)$ and stride $(16, d/8)$, is utilized to map the token sequences of both views ($\boldsymbol{X}_{tem}^{TF}$, $\boldsymbol{X}_{spe}^{TF}$) to features $\{\boldsymbol{z}^{tem}_{i},\boldsymbol{z}^{spe}_{i}\}_{i=1}^{N}$ in a shared feature space, where $\boldsymbol{z}^{tem}_{i}$ and $\boldsymbol{z}^{spe}_{i}$ are the feature vectors of the temporal view and spectral view from the $i$-th sample, respectively. The $N$ denotes the batch size. Feature similarity is assessed using the cosine distance $sim(\boldsymbol{u}, \boldsymbol{v}) = \boldsymbol{u}^{T}\boldsymbol{v}/ \Vert \boldsymbol{u} \Vert_2 \Vert \boldsymbol{v} \Vert_2$, and the loss function with respect to positive pairs $(\boldsymbol{z}^{tem}_{i},\boldsymbol{z}^{spe}_{i})$ is defined as:

    \begin{equation}
        g(\boldsymbol{z}^{tem}_{i},\boldsymbol{z}^{spe}_{i})  = \frac{e^{sim(\boldsymbol{z}^{tem}_{i},\boldsymbol{z}^{spe}_{i})/\tau}}{\sum_{\substack{j=1 \\ j\neq i}}^{N} 
        e^{sim(\boldsymbol{z}^{tem}_{i},\boldsymbol{z}^{tem}_{j})/\tau} + \sum_{j=1}^{N} e^{sim(\boldsymbol{z}^{tem}_{i},\boldsymbol{z}^{spe}_{j})/\tau}},
    \end{equation}
    
    \begin{equation}
        \mathcal{L}_{i}  = \log \left[ g(\boldsymbol{z}^{tem}_{i},\boldsymbol{z}^{spe}_{i})\right] + \log\left[ g(\boldsymbol{z}^{spe}_{i},\boldsymbol{z}^{tem}_{i})\right],
    \end{equation}
    where $\tau$ is a temporal parameter to adjust the scale. Then the multi-view consistency loss can be formulated as follows:

    \begin{equation}
        \mathcal{L}_{multi\mbox{-}view} = - \dfrac{1}{2N} \sum_{i=1}^{N} \mathcal{L}_{i}.
    \end{equation}  

    \subsection{Fusion Module}	
    Although temporal tokens and spectral tokens are in different domains, they are naturally aligned along the time dimension. To effectively extract high-level discriminative features of EEG signals, we design an attention-based fusion module, which fuses the temporal and spectral features by taking advantage of the alignment between the two view features and the time-series dependencies of the fused features.
    
    The fusion module is proposed after the cross-view interaction module to fuse temporal and spectral tokens, thereby generating comprehensive fusion features for classification (see Fig. \ref{model framework}). Since temporal tokens and spectral tokens are extracted from EEG temporal signals and spectrogram images during the same time period, they naturally align in the time dimension. As shown in Fig. \ref{model framework}. Initially, we token-wise concatenate the two-view tokens and linearly project them into a $d$-dimensional feature space, each of them containing temporal-spectral information for its corresponding time period. Then we designate these tokens as queries and employ both temporal and spectral tokens as keys and values in a cross-attention layer to obtain fusion tokens ($\boldsymbol{X}_{fus}\in \mathbb{R}^{n\times d}$). At the end of the fusion module, we design a convolution layer with 16 convolution kernels to aggregate all fusion tokens and flatten the output feature into a feature vector as the fusion feature ($\boldsymbol{z}_{fus}$) for final prediction. The convolution kernel size is $(16, d/8)$ with a stride of $(16,d/8)$. 

    \subsection{Subject-Specific Adapter}	
    Owing to variations in data distribution among different subjects, the pre-trained model needs to transfer the learned general patterns to the data distribution of new subjects for efficient decoding. Moreover, to reduce the preparation time of new subject training data acquisition as much as possible, the amount of new subject data that can be used for fine-tuning the pre-training model is limited. The adapter-based fine-tuning can realize the transfer of pre-trained models to downstream tasks by updating a few parameters, which can avoid overfitting caused by limited data \cite{houlsby2019parameter}. Thus, we introduce a simple and effective Subject-specific Adapter (SA) to rapidly transfer the knowledge of the pre-trained model learned from existing subjects’ data to data characteristics of new subjects using a small amount of data from the new subject.
    
    To simplify the model, we only add an SA to the last layer of the fusion module, and the structure of the adapter is the same as the last convolution layer in the fusion module. The adapter aggregates all fusion tokens and flattens the output features into vectors as subject-specific features ($\boldsymbol{z}_{sub}$), which are subsequently concatenated with fusion features as $[\boldsymbol{z}_{fus}||\boldsymbol{z}_{sub}]$ for the final prediction, where $[\boldsymbol{x}||\boldsymbol{y}]$ represents concatenating $\boldsymbol{x}$ and $\boldsymbol{y}$ to a new vector.

    In the pre-training stage, the TSformer is optimized using the overall loss function which consists of two parts: the multi-view consistency loss and the classification loss. The classification loss is based on cross-entropy loss:
    \begin{equation}    
	   \mathcal{L}_{cross\mbox{-}entropy} = -\dfrac{1}{N}\sum_{n=1}^{N}\sum_{r=1}^{R}y_{n,r}\log \hat{y}_{n,r},
    \end{equation}
    where $R$ denotes the number of classes, with $y$ indicating the real label and $\hat{y}$ representing the value calculated by a linear layer based on the fusion features ($\boldsymbol{z}_{fus}$). The overall loss function is formulated as follows:
    \begin{equation}    
	   \mathcal{L}_{overall} = \mathcal{L}_{cross\mbox{-}entropy} + \mathcal{L}_{multi\mbox{-}view}.
    \end{equation}

    In the fine-tuning stage, we only employ the cross-entropy loss ($\mathcal{L}_{cross\mbox{-}
    entropy}$) to update the parameters of the SA and the parameters used for projecting the subject-specific features within the linear layer. The value $\hat{y}$ in the cross-entropy loss is calculated by a linear layer using the concatenated features ($[\boldsymbol{z}_{fus}||\boldsymbol{z}_{sub}]$). Through fine-tuning the SA, TSformer-SA can be trained with a few data of new subjects without modifying all the pre-trained parameters in TSformer. This approach allows the model to preserve its capability to extract task-related features obtained in the pre-training stage while also facilitating rapid adaptation to data distribution of new subjects for efficient decoding.

    \section{Experiments}
    \subsection{Hyperparameters Setting and Implementation Details}
    The symbols and values of network hyperparameters are summarized in Table \ref{hyperparameters}. We employ the Mexican hat wavelet for the CWT with a scale set to 20, as P300 detection emphasizes low-frequency EEG information. To prevent overfitting, we appropriately reduce the size of the model by setting $d$ as 128 and $h$ as 4. 

    \setlength{\tabcolsep}{2mm}{
	\begin{table}[htbp]
            \footnotesize
            \centering
            \renewcommand\arraystretch{1.25}
            \caption{The symbols and values summary of network hyperparameters.}
            \label{hyperparameters}
            \begin{tabular}{cll}
                \toprule[1.2pt]
                \textbf{Symbol} & \textbf{Definition}  & \textbf{Value} \\
                \hline
                $C$ & Number of EEG channels  & 64 \\
                $T$ & Number of EEG sampling points & 250 \\
                $S$ & Size of scale in CWT  & 20  \\
                $t$ & Slice length in the slice embedding layer  & 5 \\
                $n$ & Length of token sequence & 50 \\
                $d$ & Embedding dimension of all the tokens & 128 \\
                $h$ & Number of heads  & 4 \\
                $\tau$ & Temporal parameter of multi-view consistency loss & 0.2 \\
                \bottomrule[1.2pt] 
            \end{tabular}
	\end{table}} 

    In each of the three tasks, each subject will be treated as the test subject in turn, which can be considered as the new subject. The remaining subjects constitute the training subject, which can be considered as the existing subjects. The TSformer-SA is initially trained on data from existing subjects in the pre-training stage using the $\mathcal{L}_{overall}$. Then, the data from the new subject are used as the training set to train the subject-specific adapter in the fine-tuning stage using the $\mathcal{L}_{cross\mbox{-}entropy}$.
 
    We implement our model within the PyTorch framework. In both training stages, we utilize the Adam optimizer \cite{kingma2014adam} for model optimization with a learning rate set to 0.0005, which is reduced by 20\% every 40 epochs. Additionally, the L2 regularization is adopted with a weight decay coefficient of 0.01. The batch size ($N$) is set to 256 and the training process is limited to a maximum of 100 epochs in the pre-training stage and 50 epochs in the fine-tuning stage.

    \begin{figure*}[!htbp]
		\centering
        
		\subfigure[Subject-dependent decoding]{
			\includegraphics[width=0.31\linewidth]{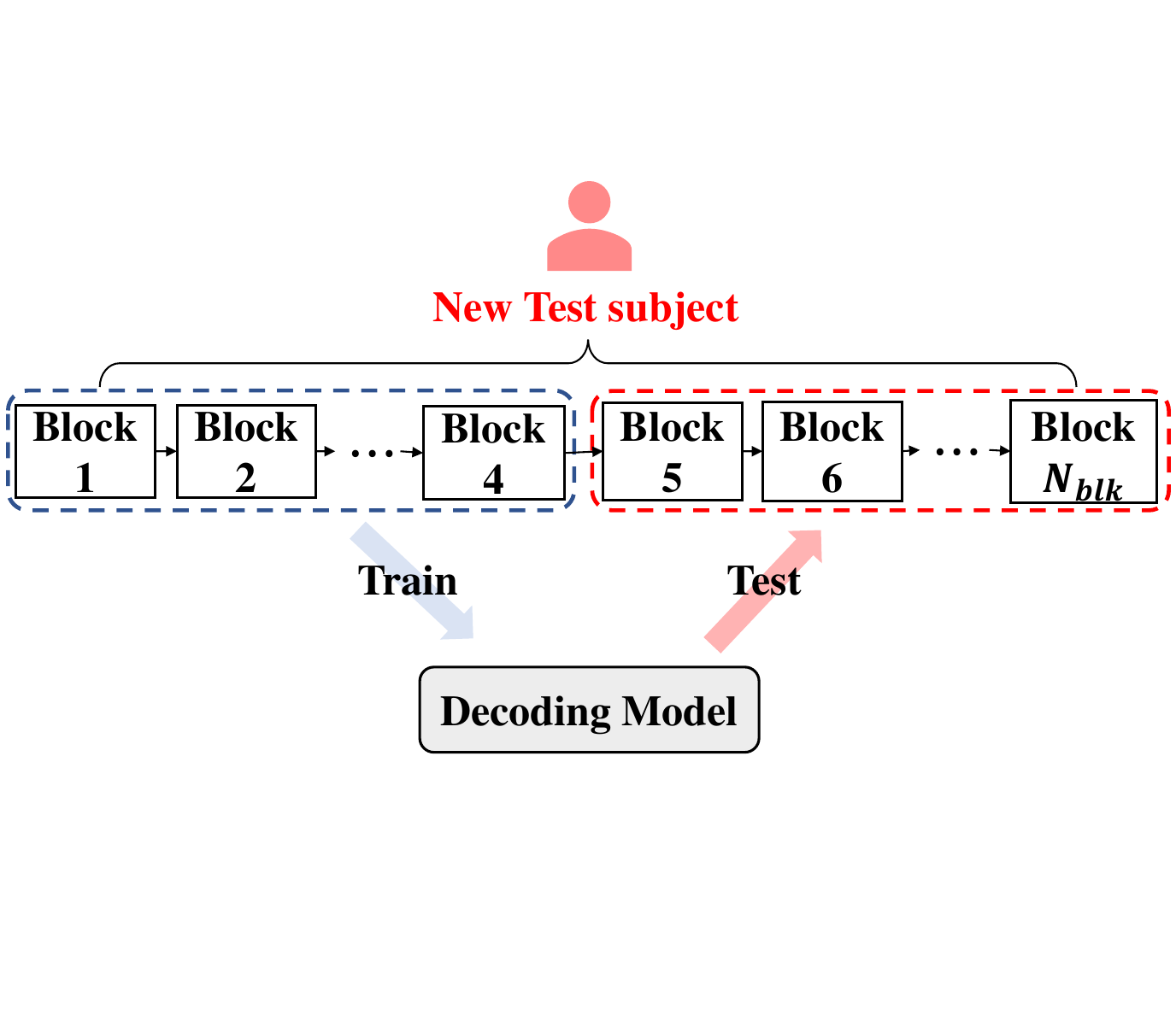}
		}
		\subfigure[Subject-independent decoding]{
			\includegraphics[width=0.31\linewidth]{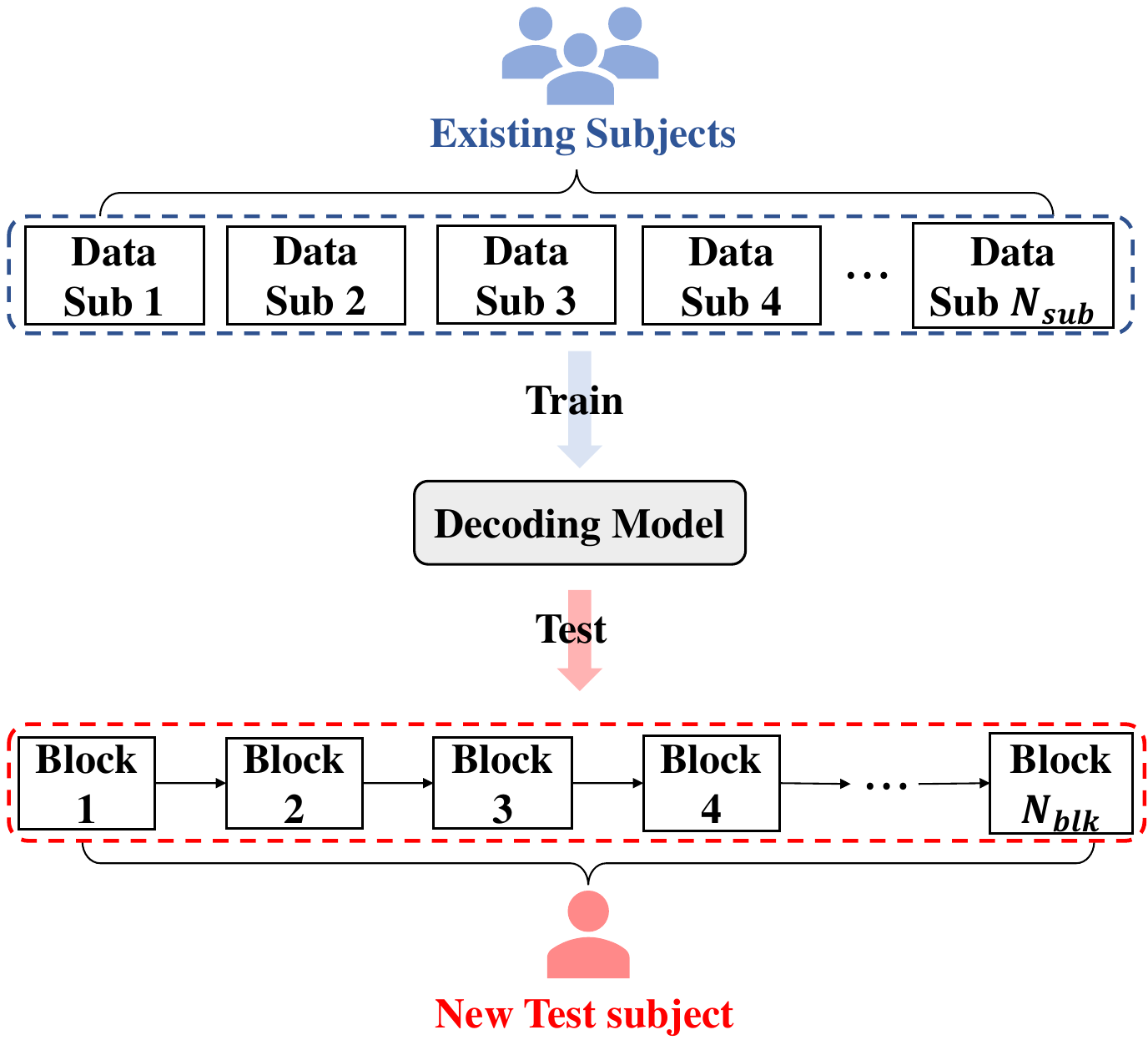}
		}
		\subfigure[Two-stage training]{
			\includegraphics[width=0.31\linewidth]{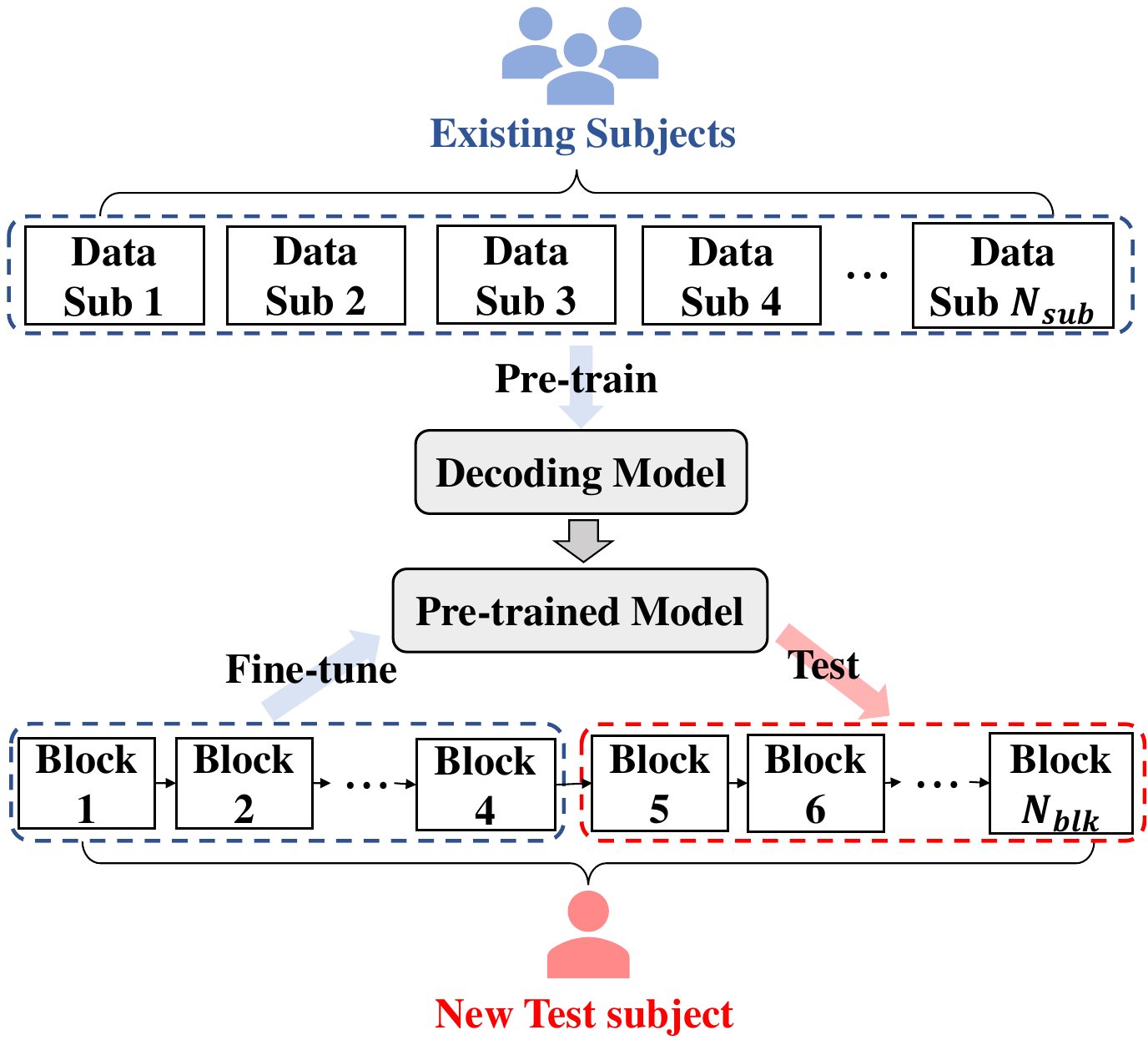}
		}
		\caption{The experimental flow charts of (a) subject-dependent decoding, (b) subject-independent decoding, and (c) two-stage training.}
        \label{experimental setting}
    \end{figure*}

    \subsection{Comparison Methods}
    We compare our proposed model with two conventional machine learning methods, five CNN-based methods, and three Transformer-based methods. 
    
    For fair comparisons, we resample our EEG signals as described in the comparison methods and modify the channel number to the same as ours. We re-implement the model architecture of the comparison methods exactly and optimize these models using the same optimization techniques and parameter settings as described in the source literature. The comparison methods are as follows:

    \subsubsection{Conventional machine learning methods}
    \medskip
    \begin{itemize}
        \item HDCA \cite{HDCA}: a linear discrimination method that learns weights on the channel and time window of EEG. We set the time window to 25 sampling points.
        \item MDRM \cite{MDRM}: a Riemannian geometry classifier, classifies samples according to the geodesic distance to the center of the category. The method is re-implemented based on the Python package pyriemann.
    \end{itemize}

    \subsubsection{CNN-based methods}
    \medskip
    \begin{itemize}
        \item MCNN \cite{MCNN}: a three-layer CNN proposed by Manor, we re-implement the network under the description in \cite{MCNN}.
        \item EEGNet \cite{EEGNet}: a CNN-based method with depthwise and separable convolution layers. The implementation refers to https://github.com/vlawhern/arl-eegmodels.
        \item LeeNet \cite{LeeCNN}: a convolutional neural network based on EEGNet, which is suitable for large ERP dataset training. The method is implemented according to \cite{LeeCNN}.
        \item PLNet \cite{PLNet}: a CNN-based method using the phase-locked characteristic of ERP signals to extract spatiotemporal features. The method is implemented according to \cite{PLNet}.
        \item PPNN \cite{PPNN}: a CNN-based method with dilated temporal convolution layers to preserve phase information. The implementation refers to \cite{PPNN}.
    \end{itemize}

    \subsubsection{Transformer-based methods}
    \medskip
    \begin{itemize}
        \item TCN-T \cite{TCN-T}: a network architecture based on Temporal Convolution Network and Transformer. The implementation refers to \cite{TCN-T}.
        \item EEG-conformer \cite{song2022eeg}: a compact convolutional Transformer, which encapsulates local and global features in a unified EEG classification framework. The implementation refers to https://github.com/eeyhsong/EEG-Conformer.
        \item CAW-STST \cite{luo2023dual}: a dual-branch spatio-temporal-spectral Transformer with a channel attention weighting module. The implementation refers to https://github.com/ljbuaa/VisualDecoding.
    \end{itemize}
    
    \subsection{Experimental Setup}
    In section 3, we conduct three distinct RSVP experiments for three different application scenarios. The subjects in the three experiments are not repeated. Consequently, the Task plane, Task car, and Task people can be viewed as three independent RSVP datasets. When evaluating the model's performance, we conduct all experiments on three datasets separately.

    To assess our model's performance, we first compare it with subject-dependent decoding methods. Next, we evaluate TSformer's capability to extract task-related features from existing subjects and generalize them to new subjects by comparing the subject-independent decoding performance of TSformer and comparison methods. To validate the effectiveness of the proposed subject-specific adapter, we compare our method with comparison methods using the same two-stage training setting. The process of these three experimental settings are as follows:
    \begin{itemize}[leftmargin=0cm, itemindent=0.5cm]
        \item[1)] Subject-dependent decoding (see Fig \ref{experimental setting} (a)) requires the training set and test set from the same subject collected in the same RSVP task. In each of the three tasks, for each subject, the data of the first $b$ ($b=1, 2, 3, 4$) blocks are designated as the training set, while the remaining blocks are used as the test set. This setting simulates the practical use of RSVP-BCI systems.
        \item[2)] Subject-independent decoding (see Fig \ref{experimental setting} (b)) requires the training set and test set from different subjects collected in the same RSVP task. In each of the three tasks, the experiments are conducted using the Leave-One-Subject-Out (LOSO) settings. In each task, each subject will be treated as the test set in turn and the remaining subjects constitute the training set. 
        \item[3)] Two-stage training strategy (see Fig \ref{experimental setting} (c)) involves the pre-training stage and fine-tuning stage. For each task, every subject takes a turn as the test subject. First, the model is pre-trained using the data from other subjects in the same RSVP task in the pre-training stage. Then, the pre-trained model is fine-tuned using the data from the first b blocks of the test subjects. Finally, the model's performance is evaluated on the remaining data from the test subjects. 
    \end{itemize}
    
    Moreover, to overcome the impact of extreme class imbalance in RSVP tasks, we employ the resampling technique. Specifically, we down-sample the nontarget class to match the sample size of the target class. This down-sampling procedure is only employed in the training set.
	
    \subsection{Evaluation Metrics}
    We evaluate model performance using Balanced Accuracy (BA), True Positive Rate (TPR), and False Positive Rate (FPR). All experiments are conducted separately on each of the three datasets, so the average of BA, TPR, and FPR for each test subject in the three tasks are calculated separately. The calculation formulas are as follows:

    \begin{equation}
    	\left\{ 
        \begin{aligned}
        	BA &=  \left( \frac{TP}{TP+FN}+\frac{TN}{TN+FP} \right)/2, \\[0.2cm]
        	TPR &= \frac{TP}{TP+FN}, \\[0.2cm]
        	FPR &= \frac{FP}{TN+FP}.
        \end{aligned} \right.
    \end{equation}
    where TP stands for the number of correctly classified positive samples, FN represents the number of incorrectly classified positive samples, TN denotes the number of correctly classified negative samples, and FP signifies the number of incorrectly classified negative samples.

    \setlength{\tabcolsep}{2.8mm}{
	\begin{table*}[htbp!]
        \footnotesize
        \centering
        \renewcommand\arraystretch{1.3}
        \caption{Comparisons of BA, TPR, and FPR of TSformer-SA and compared methods in subject-dependent decoding on three test tasks (mean).}
        \label{subject-dependent decoding}
        \begin{tabular}{llllllllll}
            \toprule[1.2pt]
            \multirow{3}{*}{\textbf{ Methods}}& \multicolumn{3}{c}{\multirow{2}{*}{\textbf{Task plane}}} & \multicolumn{3}{c}{\multirow{2}{*}{\textbf{ Task car}}} & \multicolumn{3}{c}{\multirow{2}{*}{\textbf{ Task people}}}\\
            & \multicolumn{3}{c}{} & \multicolumn{3}{c}{} & \multicolumn{3}{c}{}\\
            \cmidrule(lr){2-4}\cmidrule(lr){5-7}\cmidrule(lr){8-10}
            & \textbf{BA $(\%)$}$\uparrow$ & \textbf{TPR $(\%)$ $\uparrow$} & \textbf{FPR $(\%)$}$\downarrow$ & \textbf{BA $(\%)$}$\uparrow$ & \textbf{TPR $(\%)$ $\uparrow$} & \textbf{FPR $(\%)$}$\downarrow$ & \textbf{BA $(\%)$}$\uparrow$ & \textbf{TPR $(\%)$ $\uparrow$} & \textbf{FPR $(\%)$}$\downarrow$ \\
            \hline
            HDCA&  $\underline{88.05}  ^{\star\star}$ & $86.16 ^{\star\star}$& $ \underline{10.05}^{\star}$&   $\underline{85.82} ^{\star\star}$ & $\underline{84.33} ^{\star\star}$& $\underline{12.68} ^{\star\star}$&  $ 88.31^{\star\star\star}$ & $87.03 ^{\star\star}$& $\underline{10.41} ^{\star}$ \\
            MDRM &  $ 82.62 ^{\star\star\star}$ & $80.48 ^{\star\star\star}$& $ 15.23^{\star\star}$&   $83.89 ^{\star\star\star}$ & $ 81.08^{\star\star\star}$& $13.31 ^{\star\star\star}$&  $ 84.97^{\star\star\star}$ & $80.52 ^{\star\star\star}$& $10.57 ^{\star\star}$ \\
            \midrule
            MCNN &  $86.49  ^{\star\star\star}$ & $\underline{86.98} ^{\star\star}$& $ 14.01  ^{\star\star}$&   $84.28 ^{\star\star\star}$ & $83.24 ^{\star\star\star}$& $14.67 ^{\star\star\star}$&  $ 85.89^{\star\star\star}$ & $85.10 ^{\star\star\star}$& $13.32 ^{\star\star\star}$ \\
            EEGNet&  $84.57  ^{\star\star\star}$ & $82.78 ^{\star\star\star}$& $13.64 ^{\star\star}$&   $83.02 ^{\star\star\star}$ & $ 80.51^{\star\star\star}$& $14.46 ^{\star\star\star}$&  $ 86.87^{\star\star\star}$ & $84.79 ^{\star\star\star}$& $11.06 ^{\star\star\star}$ \\
            LeeNet&  $85.19  ^{\star\star\star}$ & $83.84 ^{\star\star\star}$& $13.47 ^{\star\star}$&   $83.39 ^{\star\star\star}$ & $ 80.01^{\star\star\star}$& $13.23 ^{\star\star\star}$&  $ 86.99^{\star\star\star}$ & $85.02 ^{\star\star\star}$& $11.04 ^{\star\star\star}$ \\
            PLNet&  $81.50  ^{\star\star\star}$ & $80.46 ^{\star\star\star}$& $ 17.45^{\star\star}$&   $79.74 ^{\star\star\star}$ & $ 80.21^{\star\star\star}$& $20.73 ^{\star\star\star}$&  $83.75 ^{\star\star\star}$ & $82.85 ^{\star\star\star}$& $15.36 ^{\star\star\star}$ \\
            PPNN&  $85.88  ^{\star\star\star}$ & $85.77 ^{\star\star\star}$& $ 14.02^{\star\star}$&   $84.53 ^{\star\star\star}$ & $ 83.74^{\star\star\star}$& $14.68 ^{\star\star\star}$&  $87.02 ^{\star\star\star}$ & $85.62 ^{\star\star\star}$& $11.59 ^{\star\star\star}$ \\
            \midrule
            TCN-T&  $87.85  ^{\star\star}$ & $85.89 ^{\star\star\star}$& $ 10.19^{\star}$&   $84.25 ^{\star\star\star}$ & $82.88 ^{\star\star\star}$& $14.38 ^{\star\star\star}$&  $85.26 ^{\star\star\star}$ & $81.80 ^{\star\star\star}$& $11.28 ^{\star\star\star}$ \\
            EEG-conformer&  $87.75  ^{\star\star\star}$ & $86.08 ^{\star\star\star}$& $10.58 ^{\star}$&   $83.83 ^{\star\star\star}$ & $81.94 ^{\star\star\star}$& $14.27 ^{\star\star\star}$&  $\underline{88.41} ^{\star\star}$ & $\underline{87.92} ^{\star}$& $11.11 ^{\star\star\star}$ \\
            CAW-STST&  $81.22 ^{\star\star\star}$ & $75.17 ^{\star\star\star}$& $12.72 ^{\star\star}$&   $83.06 ^{\star\star\star}$ & $ 83.58^{\star\star\star}$& $17.45 ^{\star\star\star}$&  $80.51 ^{\star\star\star}$ & $85.35 ^{\star\star\star}$& $24.33 ^{\star\star\star}$ \\
            \midrule
            TSformer-SA&  $\mathbf{90.29}$ & $\mathbf{89.21}$& $\mathbf{8.64} $&   $\mathbf{88.42} $ & $\mathbf{87.24}$& $\mathbf{10.39} $&  $\mathbf{90.20} $ & $\mathbf{89.36}$& $\mathbf{8.97} $ \\
            \bottomrule[1.2pt] 
        \end{tabular}
        \begin{tablenotes}
            \item \footnotesize The asterisks in the table indicate a significant difference between TSformer-SA and the comparison methods by paired t-tests ($^{\star}$ $p<0.05$, $^{\star\star}$ $p<0.01$, $^{\star\star\star}$ $p<0.001$). The best results are highlighted in bold and the second-best results are denoted by underline.
        \end{tablenotes}
    \end{table*}}

    \subsection{Statistical Analysis}
    This study employs one-way and two-way repeated measures analysis of variance (ANOVA) to assess the impact of various factors on RSVP decoding performance. The Greenhouse-Geisser correction is applied if the data do not conform to the Sphericity assumption. Post-hoc analyses are conducted for each significant factor, and the Holm-Bonferroni correction is applied to adjust the p-values in pairwise comparisons. The significance level is set to 0.05. 

    \section{Results and Discussion}	

    \subsection{Subject-Dependent Experiments}
    The comparison experiment is conducted to compare the classification performance of TSformer-SA and comparison methods in subject-dependent decoding. The results of three RSVP tasks are summarized in Table \ref{subject-dependent decoding}.
    
    The one-way repeated measures ANOVA reveals a significant main effect of the method on the three evaluation metrics including BA, TPR, and FPR in all three tasks (all tasks: BA, $p<0.01$; TPR, $p<0.01$; FPR, $p<0.001$). Post-hoc tests indicate that the BA of our proposed TSformer-SA is significantly superior to that of all comparison methods in three tasks (all tasks: $p<0.01$). Specifically, TSformer-SA significantly outperforms comparison methods on both TPR and FPR in all three tasks (all tasks: TPR, $p<0.05$; FPR, $p<0.05$). These results demonstrate the effectiveness of our method in enhancing RSVP decoding performance.

    Among conventional machine learning methods, HDCA significantly outperforms MDRM on BA and TPR (all tasks: BA, $p<0.01$; TPR, $p<0.05$), while the FPR of HDCA is lower than that of MDRM in three tasks (Task plane and Task car: $p<0.05$; Task people: $p=0.79$). For deep learning comparison methods including CNN-based and Transformer-based methods, TCN-T, PPNN, and EEG-conformer separately achieve the highest BA in Task plane, Task car, and Task people. However, the BA of HDCA is significantly higher than that of TCN-T in the Task plane ($p<0.001$) and PPNN in the Task car ($p<0.001$), and there is no statistical difference between the BA of HDCA and EEG-conformer in Task people ($p=0.85$). This may be attributed to that deep learning methods require more training data to avoid overfitting and achieve better performance compared to conventional machine learning methods. However, due to the limitation of preparation time and subject status, it is difficult to collect sufficient data from new subjects for deep learning model training. Consequently, the deep learning methods cannot be sufficiently optimized with the constraints of limited training data, resulting in inferior performance compared to HDCA. In contrast, our proposed TSformer-SA significantly outperforms HDCA on BA, TPR, and FPR (all tasks: BA, $p<0.01$; TPR, $p<0.01$; FPR, $p<0.05$). These results indicate that TSformer-SA can achieve superior decoding performance under limited training data from new subjects. 
    
    Similar to TSformer-SA, CAW-STST also employs a multi-view learning framework, but it achieves the lowest BA in both Task plane and Task people. Thereinto, the BA of CAW-STST is significantly lower than the other comparison methods in the Task plane ($p<0.05$), except for PLNet ($p=0.43$). In the Task people, the BA of CAW-STST is also significantly lower than the other comparison methods ($p<0.01$). This could be attributed to the excessive number of network parameters trained from scratch and the limited training data, which leads to overfitting during the training process. In contrast, TSformer-SA is initially trained on data from existing subjects to extract comprehensive task-related representations from diverse subjects effectively. Subsequently, the learned general patterns are extended to the data distribution of new subjects by fine-tuning the subject-specific adapter with a few parameters, which prevents the risk of overfitting caused by limited training data. Thus, the TSformer-SA achieves superior decoding performance while reducing the reliance on training data from new subjects.

    \setlength{\tabcolsep}{2.8mm}{
	\begin{table*}[htbp!]
            \footnotesize
            \renewcommand\arraystretch{1.3}
            \centering
            \caption{Comparisons of BA, TPR, and FPR of TSformer and compared methods in subject-independent decoding on three test tasks (mean).}
            \label{subject-independent decoding}
            \begin{tabular}{llllllllll}
                \toprule[1.2pt]
                \multirow{3}{*}{\textbf{ Methods}}& \multicolumn{3}{c}{\multirow{2}{*}{\textbf{ Task plane}}} & \multicolumn{3}{c}{\multirow{2}{*}{\textbf{ Task car}}} & \multicolumn{3}{c}{\multirow{2}{*}{\textbf{ Task people}}}\\
                & \multicolumn{3}{c}{} & \multicolumn{3}{c}{} & \multicolumn{3}{c}{}\\
                \cmidrule(lr){2-4}\cmidrule(lr){5-7}\cmidrule(lr){8-10}
                & \textbf{BA $(\%)$}$\uparrow$ & \textbf{TPR $(\%)$ $\uparrow$} & \textbf{FPR $(\%)$}$\downarrow$ & \textbf{BA $(\%)$}$\uparrow$ & \textbf{TPR $(\%)$ $\uparrow$} & \textbf{FPR $(\%)$}$\downarrow$ & \textbf{BA $(\%)$}$\uparrow$ & \textbf{TPR $(\%)$ $\uparrow$} & \textbf{FPR $(\%)$}$\downarrow$ \\
                \hline
                HDCA&  $81.37^{\star\star\star}$ & $ 78.34 ^{\star\star\star}$& $15.59^{\star\star\star}$&   $80.12^{\star\star\star}$ & $ 78.40^{\star\star}$& $18.14^{\star\star\star}$&  $82.60^{\star\star\star}$ & $80.48^{\star\star\star}$& $15.29^{\star\star\star}$ \\
                MDRM & $82.19^{\star\star\star}$ & $76.10^{\star\star\star}$& $11.73^{\star\star\star}$& $80.92^{\star\star\star}$ & $ 77.49 ^{\star\star}$& $15.64^{\star\star\star}$& $82.69^{\star\star\star}$ & $78.63^{\star\star\star}$& $13.25^{\star\star\star}$ \\
                \midrule
                MCNN & $83.57^{\star\star\star}$ & $76.93^{\star\star\star}$& $9.77^{\star\star}$ &$83.95^{\star\star}$ & $ 79.77^{\star\star}$& $11.86^{\star}$& $83.05^{\star\star\star}$ & $79.16^{\star\star\star}$&$13.05^{\star\star\star}$   \\
                EEGNet& $84.21^{\star\star\star}$ & $ 81.01^{\star\star\star}$& $12.59^{\star\star\star}$& $82.85^{\star\star\star}$ & $\underline{80.75} ^{\star}$& $15.04^{\star\star\star}$& $79.99^{\star\star\star}$& $78.88^{\star\star\star}$& $18.89^{\star\star\star}$    \\
                LeeNet& $84.91^{\star\star\star}$ & $ 81.59 ^{\star\star}$& $11.78^{\star\star\star}$& $83.01^{\star\star\star}$ & $ 79.21^{\star\star}$& $13.18^{\star\star}$& $85.03^{\star\star\star}$& $80.60^{\star\star\star}$& $10.54^{\star\star\star}$ \\
                PLNet& $79.78^{\star\star\star}$ & $73.81^{\star\star\star}$& $14.24^{\star\star\star}$& $78.11^{\star\star\star}$ & $ 73.34^{\star\star\star}$& $17.10^{\star\star\star}$& $79.38^{\star\star\star}$& $78.21^{\star\star\star}$ & $19.45^{\star\star\star}$   \\
                PPNN& $86.26^{\star\star\star}$ &  $81.49^{\star\star\star}$&  $8.97^{\star}$&  $83.42^{\star\star}$ & $78.36^{\star\star}$& $11.52$ & $85.48^{\star\star\star}$& $83.80^{\star\star}$& $12.83^{\star\star\star}$   \\
                \midrule
                TCN-T& $85.92^{\star\star\star}$ & $80.19^{\star\star\star}$& $\underline{8.34}$& $\underline{84.50}^{\star}$ & $ 80.02^{\star\star}$& $\underline{11.01}$& $86.68^{\star\star}$& $83.51^{\star\star}$& $\underline{10.16}$   \\
                EEG-conformer& $85.51^{\star\star\star}$ & $81.09^{\star\star}$& $10.06^{\star\star}$& $ 83.56^{\star\star}$ & $80.71^{\star\star}$& $ 13.58^{\star\star}$& $86.14^{\star\star\star}$ & $ 83.66^{\star\star}$& $11.38^{\star\star\star}$    \\
                CAW-STST& $\underline{86.62}^{\star\star}$ & $\underline{81.72}^{\star\star}$& $8.47$& $ 84.10^{\star\star}$ & $ 80.44^{\star}$& $12.22^{\star}$& $\underline{86.82}^{\star\star\star}$ & $\underline{84.65}^{\star\star}$& $10.99^{\star\star}$    \\
                \midrule
                TSformer & $\mathbf{88.01}$ & $\mathbf{83.88}$ & $\mathbf{7.86}$ & $\mathbf{85.88}$  & $\mathbf{82.69}$ & $\mathbf{10.92}$ & $\mathbf{ 88.08}$  & $\mathbf{85.91}$ & $\mathbf{9.74}$  \\
                \bottomrule[1.2pt] 
            \end{tabular}
            \begin{tablenotes}
                \item \footnotesize The asterisks in the table indicate a significant difference between TSformer and the comparison methods by paired t-tests ($^{\star}$ $p<0.05$, $^{\star\star}$ $p<0.01$, $^{\star\star\star}$ $p<0.001$). The best results are highlighted in bold and the second-best results are denoted by underline.
            \end{tablenotes}
    \end{table*}}
    
    \subsection{Subject-Independent Experiments}
    To evaluate the ability of TSformer to extract task-related features from various subjects obtained in the pre-training stage, we conduct a subject-independent comparison experiment. The TSformer and the compared methods are trained on the training subjects and then directly tested on the new test subject.

    Table \ref{subject-independent decoding} presents the classification performance in three RSVP tasks. The one-way repeated measures ANOVA in each task demonstrates a significant main effect of the method on BA, TPR, and FPR (all tasks: BA, $p<0.001$; TPR, $p<0.01$; FPR, $p<0.001$). Subsequent post-hoc tests show that our proposed TSformer significantly outperforms all comparison methods on BA and TPR (all tasks: BA, $p<0.05$; TPR, $p<0.05$), and the FPR of TSformer is lower than that of all the comparison methods in three tasks. These results illustrate the capacity of our method to effectively extract task-related features from diverse subjects' data for RSVP decoding.
 
    Among all the comparison methods, TCN-T achieves the highest BA in the Task car and CAW-STST achieves the highest BA in the Task plane and people among all the comparison methods. Compared to TCN-T which focuses on the temporal characteristics, TSformer leverages both temporal and spectral information of EEG signals to extract more comprehensive representations for classification, which achieves significantly higher BA and TPR in all three tasks (all tasks: BA, $p<0.05$; TPR, $p<0.01$). Moreover, CAW-STST utilizes the same multi-view information of EEG signals as TSformer, but TSformer achieves significantly higher BA and TPR compared to CAW-STST (all tasks: BA, $p<0.01$; TPR, $p<0.05$) and exhibits a lower FPR than CAW-STST (Task plane: $p=0.17$; Task car and people: $p<0.05$). This indicates that our proposed multi-view learning framework obtains more discriminative high-level fusion features and improves the RSVP decoding performance, where the cross-view interaction module and fusion module in TSformer effectively enhance complementary information transfer and common feature extraction between the two views. 
    
    HDCA and MDRM exhibit significantly lower BA compared to five deep learning comparison methods including LeeNet, PPNN, TCN-T, EEG-conformer, and CAW-STST in all three tasks (all tasks: $p<0.01$). This contrasts with the results of subject-dependent decoding experiment in Table \ref{subject-dependent decoding}, where the conventional machine learning method HDCA achieves the highest BA in two tasks. These results can be attributed to two aspects. Firstly, conventional machine learning methods including HDCA and MDRM are limited in their capacity to effectively model complex data distributions and extract task-related features across different subjects, which limits their generalization performance. Secondly, sufficient training data from various subjects allows decoding methods based on deep learning to be efficiently optimized, which enables them to learn task-related representation extraction across diverse subjects and improve generalization performance for decoding data from new subjects. 

    \setlength{\tabcolsep}{2.0mm}{
	\begin{table*}[htbp]
            \footnotesize
            \renewcommand\arraystretch{1.3}
            \centering
            \caption{Ablation study of TSformer-SA in subject-dependent decoding on three test tasks (mean).}
            \label{ablation study subject-dependent decoding}
            \begin{tabular}{ccccclllllllll}
                \toprule[1.2pt]
                \multirow{3}{*}{\textbf{Methods}}&\multicolumn{4}{c}{\multirow{2}{*}{\textbf{Modules}}}& \multicolumn{3}{c}{\multirow{2}{*}{\textbf{Task plane}}} & \multicolumn{3}{c}{\multirow{2}{*}{\textbf{Task car}}} & \multicolumn{3}{c}{\multirow{2}{*}{\textbf{Task people}}}\\
                \multicolumn{4}{c}{}& \multicolumn{3}{c}{} & \multicolumn{3}{c}{} & \multicolumn{3}{c}{}\\
                \cmidrule(lr){2-5}\cmidrule(lr){6-8}\cmidrule(lr){9-11}\cmidrule(lr){12-14}
                &\textbf{W} & \textbf{T} & \textbf{F} & \textbf{C} & \textbf{BA $(\%)\uparrow$} & \textbf{TPR $(\%)\uparrow$} & \textbf{FPR $(\%)$}$\downarrow$ & \textbf{BA $(\%)$}$\uparrow$ & \textbf{TPR $(\%)\uparrow$} & \textbf{FPR $(\%)$}$\downarrow$ & \textbf{BA $(\%)$}$\uparrow$ & \textbf{TPR $(\%)\uparrow$} & \textbf{FPR $(\%)$}$\downarrow$ \\
                \hline
                $\textbf{M1}$& $\surd$ &$-$ &$-$ &$-$ & $82.46 ^{\star\star\star}$& $ 80.53^{\star\star\star}$ & $15.60 ^{\star\star\star}$& $ 78.38^{\star\star\star}$& $ 78.07^{\star\star\star}$& $ 21.30^{\star\star\star}$& $ 
                79.21^{\star\star\star}$ & $77.15 ^{\star\star\star}$& $ 18.74^{\star\star\star}$ \\
                $\textbf{M2}$&$-$ &$\surd$ &$-$ &$-$ & $87.53 ^{\star\star\star}$& $ 86.27^{\star\star}$ & $11.21 ^{\star\star\star}$& $ 85.32^{\star\star\star}$& $83.31 ^{\star\star\star}$& $ 12.66^{\star\star\star}$& $ 
                87.48^{\star\star\star}$ & $86.71 ^{\star\star\star}$& $11.76 ^{\star\star\star}$ \\
                $\textbf{M3}$&$\surd$ &$\surd$ &$\surd$ &$-$ & $\underline{88.41} ^{\star\star\star}$& $ \underline{87.88}^{\star}$ & $\underline{11.07} ^{\star\star\star}$& $ \underline{87.41}^{\star}$& $\underline{86.03} ^{\star}$& $ \underline{11.22}$& $ 
                \underline{89.02}^{\star\star\star}$ & $\underline{88.08} ^{\star}$& $ \underline{10.03}^{\star}$ \\
                $\textbf{M4}$&$\surd$ &$\surd$ &$\surd$ &$\surd$ &  $\mathbf{90.29}$ & $\mathbf{89.21}$& $\mathbf{8.64} $&   $\mathbf{88.42} $ & $\mathbf{87.24}$& $\mathbf{10.39} $&  $\mathbf{90.20} $ & $\mathbf{89.36}$& $\mathbf{8.97} $\\
                \bottomrule[1.2pt]
            \end{tabular}
            \begin{tablenotes}
                \item \footnotesize The asterisks in the table indicate a significant difference between TSformer-SA and the ablation models by paired t-tests ($^{\star}$ $p<0.05$, $^{\star\star}$ $p<0.01$, $^{\star\star\star}$ $p<0.001$). The best results are highlighted in bold and the second-best results are denoted by underline.
                \item \footnotesize \emph{Abbreviations.} W: Wavelet view; T: Temporal view; F: Fusion module; C: Cross-view interaction module.
            \end{tablenotes}
    \end{table*}}

    \setlength{\tabcolsep}{2.0mm}{
	\begin{table*}[htbp]
            \footnotesize
            \renewcommand\arraystretch{1.3}
            \centering
            \caption{Ablation study of TSformer in subject-independent decoding on three test tasks (mean).}
            \label{ablation study subject-independent decoding}
            \begin{tabular}{ccccclllllllll}
                \toprule[1.2pt]
                \multirow{3}{*}{\textbf{ Methods}}&\multicolumn{4}{c}{\multirow{2}{*}{\textbf{ Modules}}}& \multicolumn{3}{c}{\multirow{2}{*}{\textbf{ Task plane}}} & \multicolumn{3}{c}{\multirow{2}{*}{\textbf{ Task car}}} & \multicolumn{3}{c}{\multirow{2}{*}{\textbf{ Task people}}}\\
                \multicolumn{4}{c}{}& \multicolumn{3}{c}{} & \multicolumn{3}{c}{} & \multicolumn{3}{c}{}\\
                \cmidrule(lr){2-5}\cmidrule(lr){6-8}\cmidrule(lr){9-11}\cmidrule(lr){12-14}
                &\textbf{W} & \textbf{T} & \textbf{F} & \textbf{C} & \textbf{BA $(\%)\uparrow$} & \textbf{TPR $(\%)\uparrow$} & \textbf{FPR $(\%)$}$\downarrow$ & \textbf{BA $(\%)$}$\uparrow$ & \textbf{TPR $(\%)\uparrow$} & \textbf{FPR $(\%)$}$\downarrow$ & \textbf{BA $(\%)$}$\uparrow$ & \textbf{TPR $(\%)\uparrow$} & \textbf{FPR $(\%)$}$\downarrow$ \\
                \hline
                $\textbf{M1}$&$\surd$ &$-$ &$-$ &$-$ & $78.16^{\star\star\star}$& $71.02^{\star\star\star}$ & $14.68^{\star\star\star}$& $75.30^{\star\star\star}$& $68.77^{\star\star\star}$& $18.17^{\star\star\star}$& $ 77.00^{\star\star\star}$ & $72.95^{\star\star\star}$& $18.94^{\star\star\star}$ \\
                $\textbf{M2}$&$-$ &$\surd$ &$-$ &$-$ & $85.82^{\star\star\star}$& $81.16^{\star\star}$ & $9.52^{\star\star}$& $84.11^{\star\star\star}$& $81.49^{\star}$ & $13.27^{\star\star}$& $84.78^{\star\star\star}$ &  $ 82.14^{\star\star\star}$&  $12.58^{\star\star\star}$   \\
                $\textbf{M3}$&$\surd$ &$\surd$ &$\surd$ &$-$ &   $ \underline{86.47}^{\star\star\star}$ & $\underline{82.17}^{\star\star}$& $\underline{9.22}^{\star\star}$& $ \underline{84.61}^{\star}$ & $ \underline{81.67}^{\star}$& $\underline{12.45}^{\star\star}$& $\underline{86.21}^{\star\star\star}$ &  $\underline{84.35}^{\star\star}$& $ \underline{11.93}^{\star\star\star}$    \\
                $\textbf{M4}$&$\surd$ &$\surd$ &$\surd$ &$\surd$ & $\mathbf{88.01}$ & $\mathbf{83.88 }$ & $\mathbf{7.86}$ & $\mathbf{85.88}$  & $\mathbf{82.69 }$ & $\mathbf{10.92}$ & $\mathbf{88.08}$  & $\mathbf{85.91 }$ & $\mathbf{ 9.74 }$  \\
                \bottomrule[1.2pt]
            \end{tabular}
            \begin{tablenotes}
                \item \footnotesize The asterisks in the table indicate a significant difference between TSformer and the ablation models by paired t-tests ($^{\star}$ $p<0.05$, $^{\star\star}$ $p<0.01$, $^{\star\star\star}$ $p<0.001$). The best results are highlighted in bold and the second-best results are denoted by underline.
                \item \footnotesize \emph{Abbreviations.} W: Wavelet view; T: Temporal view; F: Fusion module; C: Cross-view interaction module.
            \end{tablenotes}
    \end{table*}}
 
    \subsection{Ablation Study}
    We conduct ablation studies to evaluate the efficacy of the feature extractor, fusion module, and cross-view interaction module in TSformer-SA. Firstly, to assess the capability of the feature extractor in extracting temporal and spectral features, we only employ the feature extractor followed by a convolution layer and a linear layer for classifying spectrogram images (\textbf{M1}) and EEG temporal signals (\textbf{M2}), respectively. Then, in \textbf{M3}, we employ the feature extractor to extract both temporal and spectral features and these extracted features are then fused in the fusion module for classification. Finally, based on \textbf{M3}, we introduce the cross-view interaction module to create \textbf{M4}, which is the TSformer-SA. The results for subject-dependent decoding and subject-independent are presented in Table \ref{ablation study subject-dependent decoding} and Table \ref{ablation study subject-independent decoding}, respectively.
	
    In Table \ref{ablation study subject-dependent decoding}, one-way repeated measures ANOVA indicates a significant main effect of the proposed modules within each task on BA, TPR, and FPR (all tasks: BA, $p<0.001$; TPR, $p<0.001$; FPR, $p<0.001$). The post-hoc tests show that our proposed method (\textbf{M4}) exhibits significantly superior classification performance compared to the ablation models on BA and TPR (all tasks: BA, $p<0.05$; TPR, $p<0.05$). The FPR of (\textbf{M4}) is significantly lower than that of the ablation models in all three tasks (all tasks: $p<0.05$) except for the FPR of \textbf{M3} in Task car. However, the FPR of \textbf{M4} tends to be significantly lower than that of \textbf{M3} in  Task car ($p=0.09$). 
 
    The decoding performance of both \textbf{M1} and \textbf{M2} indicates that our proposed feature extractor can simultaneously capture the discriminative characteristics of EEG temporal signals as well as spectrogram images. After integrating the features from these two views using the fusion module, \textbf{M3} exhibits significantly improved BA (all tasks: $p<0.05$) and TPR (all tasks: $p<0.05$) compared to \textbf{M1} and \textbf{M2} in three tasks while maintaining a lower FPR in comparison to both models (Task plane: $p=0.38$; Task car and people: $p<0.01$). These results prove the effectiveness of multi-view information for enhancing decoding performance and fusion module in fusing the two view features to obtain more comprehensive fusion features. Finally, our proposed method (\textbf{M4}) incorporates cross-view interaction module into \textbf{M3}, which significantly enhances the performance on BA (all tasks: $p<0.05$), TPR (all tasks: $p<0.05$), and FPR (Task plane: $p<0.001$; Task car: $p=0.09$; Task people: $p<0.05$) when compared to \textbf{M3} in three tasks. This proves that our proposed cross-view interaction module enhances the distinction between fusion features of different classes by improving common feature extraction across two views. Similarly, the results of subject-independent decoding in Table \ref{ablation study subject-independent decoding} also validate the effectiveness of the feature extractor, fusion module, and cross-view interaction module. By integrating the features from temporal and spectral views with the fusion module, M3 improves the performance of models that use single-view information (\textbf{M1} and \textbf{M2}) for RSVP decoding. Additionally, our proposed method (\textbf{M4}) significantly outperforms \textbf{M3} on BA, TPR, and FPR in all three tasks (all tasks: BA, $p<0.05$; TPR, $p<0.05$; FPR, $p<0.01$). Hence, each component proposed in TSformer-SA contributes to enhancing RSVP decoding performance.

    \subsection{Efficiency of Subject-Specific Adapter}
    We conduct an experiment to compare TSformer-SA with TSformer and other comparison methods based on pre-training and fine-tuning strategy. In this experiment, the TSformer and comparison methods are initially pre-trained using data from existing subjects. Subsequently, the last linear layers of these models are fine-tuned on the first four blocks of the new test subject. Finally, these models are tested on the remaining blocks data of the new test subject.

    \setlength{\tabcolsep}{2.8mm}{
	\begin{table*}[htbp]
            \footnotesize
            \renewcommand\arraystretch{1.3}
            \centering
            \caption{Comparisons of BA, TPR, and FPR of TSformer-SA and compared methods using two-stage training strategy on three test tasks (mean).}
            \label{two-stage decoding}
            \begin{tabular}{llllllllll}
                \toprule[1.2pt]
                \multirow{3}{*}{\textbf{ Methods}}& \multicolumn{3}{c}{\multirow{2}{*}{\textbf{ Task plane}}} & \multicolumn{3}{c}{\multirow{2}{*}{\textbf{ Task car}}} & \multicolumn{3}{c}{\multirow{2}{*}{\textbf{ Task people}}}\\
                & \multicolumn{3}{c}{} & \multicolumn{3}{c}{} & \multicolumn{3}{c}{}\\
                \cmidrule(lr){2-4}\cmidrule(lr){5-7}\cmidrule(lr){8-10}
                & \textbf{BA $(\%)$}$\uparrow$ & \textbf{TPR $(\%)$ $\uparrow$} & \textbf{FPR $(\%)$}$\downarrow$ & \textbf{BA $(\%)$}$\uparrow$ & \textbf{TPR $(\%)$ $\uparrow$} & \textbf{FPR $(\%)$}$\downarrow$ & \textbf{BA $(\%)$}$\uparrow$ & \textbf{TPR $(\%)$ $\uparrow$} & \textbf{FPR $(\%)$}$\downarrow$ \\
                \hline
                MCNN & $86.54 ^{\star\star\star}$ & $83.68 ^{\star\star\star}$& $10.60^{\star\star}$&$ 
                84.89^{\star\star\star}$ &$  81.88^{\star\star\star}$ & $ 12.11^{\star\star\star}$& $ 
                85.27^{\star\star\star}$& $82.42 ^{\star\star\star}$ & $  11.88^{\star\star\star}$   \\
                EEGNet& $87.28 ^{\star\star\star}$ & $ 86.10^{\star\star\star}$& $11.54 ^{\star\star\star}$& $84.85^{\star\star\star}$ &$ 85.28^{\star\star}$ & $ 15 .58^{\star\star\star}$& $ 
                85.87^{\star\star\star}$& $ 86.23^{\star\star\star}$ & $  14.49^{\star\star\star}$  \\
                LeeNet& $88.07 ^{\star\star\star}$ & $ 87.15^{\star\star}$& $ 
                11.01^{\star\star\star}$ &$85.83^{\star\star\star}$ & $ 85.06^{\star\star}$& $13.39 
                ^{\star\star\star}$& $86.80 ^{\star\star\star}$ & $  85.43^{\star\star\star}$&$11.83 ^{\star\star\star}$   \\
                PLNet& $80.06 ^{\star\star\star}$ & $75.34 ^{\star\star\star}$& $ 15.23^{\star\star\star}$ &$78.89 ^{\star\star\star}$ & $ 75.96^{\star\star\star}$& $18.18^{\star\star\star}$& $ 80.02^{\star\star\star}$& $72.67 ^{\star\star\star}$ & $  12.62^{\star\star\star}$   \\
                PPNN& $87.19 ^{\star\star\star}$ & $86.83 ^{\star\star}$& $ 
                12.46^{\star\star\star}$ &$85.65^{\star\star\star}$ & $ \underline{85.98}^{\star}$& $14.69 
                ^{\star\star\star}$& $86.99 ^{\star\star\star}$ & $  85.73^{\star\star\star}$&$11.74 ^{\star\star\star}$   \\
                \midrule
                TCN-T& $87.08^{\star\star\star}$ & $85.89 ^{\star\star\star}$& $11.74^{\star\star\star}$& $ 
                85.08^{\star\star\star}$ &$ 84.96 ^{\star\star}$ & $ 14.80^{\star\star\star}$& $86.89 
                ^{\star\star\star}$& $86.83 ^{\star\star}$ & $  13.05^{\star\star\star}$   \\
                EEG-conformer& $ 87.89^{\star\star\star}$ & $ 87.35^{\star}$& $ 
                11.56^{\star\star\star}$ &$84.68^{\star\star\star}$ & $83.54 ^{\star\star\star}$& $14.19 
                ^{\star\star\star}$& $86.97 ^{\star\star\star}$ & $  86.28^{\star\star\star}$&$12.33 ^{\star\star\star}$   \\
                CAW-STST& $88.45 ^{\star\star\star}$ & $ 86.96^{\star\star\star}$& $10.05^{\star}$& $ 
                86.13^{\star\star\star}$  &$85.03^{\star\star}$ & $ 12.77^{\star\star}$& $ 
                87.73^{\star\star\star}$& $86.29 ^{\star\star\star}$ & $  10.83^{\star\star\star}$   \\
                \midrule
                TSformer& $\underline{89.55} ^{\star}$ & $ \underline{88.36}^{\star}$& $\underline{9.27} ^{\star}$ &$  \underline{87.06}^{\star\star}$ & $85.36 ^{\star\star}$& $
                \underline{11.24}^{\star}$& $\underline{88.48} ^{\star\star\star}$ &$\underline{87.17}^{\star\star\star}$&$\underline{10.21}^{\star\star\star}$   \\
                TSformer-SA & $\mathbf{90.29}$ & $\mathbf{89.21}$& $\mathbf{8.64} $&   $\mathbf{88.42} $ & $\mathbf{87.24}$& $\mathbf{10.39} $&  $\mathbf{90.20} $ & $\mathbf{89.36}$& $\mathbf{8.97}$ \\
                \bottomrule[1.2pt] 
            \end{tabular}
            \begin{tablenotes}
                \item \footnotesize The asterisks in the table indicate a significant difference between TSformer-SA and the comparison methods by paired t-tests ($^{\star}$ $p<0.05$, $^{\star\star}$ $p<0.01$, $^{\star\star\star}$ $p<0.001$). The best results are highlighted in bold and the second-best results are denoted by underline.
            \end{tablenotes}
    \end{table*}}

    The results are presented in Table \ref{two-stage decoding}. The one-way repeated measures ANOVA reveals significant main effects of the method on BA, TPR, and FPR (all tasks: BA, $p<0.001$; TPR, $p<0.001$; FPR, $p<0.001$). Post-hoc tests show that TSformer-SA significantly outperforms all the comparison methods in terms of BA, TPR, and FPR (all tasks: BA, $p<0.05$; TPR, $p<0.05$; FPR, $p<0.05$). These results indicate that although the comparison methods utilize the existing subject data to decode data from new subjects, TSformer-SA still outperforms them, which demonstrates the superiority of TSformer-SA in leveraging the learned knowledge from existing subjects to enhance decoding performance on new subjects.  

    Compared to all the comparison methods, TSformer which only fine-tunes the linear layer also achieves the best performance in all three tasks on BA and FPR (all tasks: BA, $p<0.05$; FPR, $p<0.05$). The TPR of TSformer is higher than comparison methods in Task plane and people, while it exhibits no statistically significant difference in TPR compared to PPNN in Task car ($p=0.67$), which achieves the highest TPR. These results suggest that TSformer can more effectively extract task-related features from existing subjects compared to other methods and transfer the learned knowledge to decode new subjects' data. Moreover, TSformer-SA which fine-tunes the subject-specific adapter significantly outperforms TSformer with fine-tuning linear layer on BA, TPR, and FPR in all three tasks (all tasks: BA, $p<0.05$; TPR, $p<0.05$; FPR, $p<0.05$). This demonstrates the effectiveness of the subject-specific adapter, which can maintain the learned knowledge of the pre-trained model and transfer it to the specific data distribution of new subjects, thereby enhancing RSVP decoding performance.

    \subsection{Visualization}
    
    \subsubsection{EEG Signals Analysis}
\begin{figure*}[!htbp]
		\centering
        \subfigure[Oz waveform]{
			\includegraphics[width=0.46\linewidth]{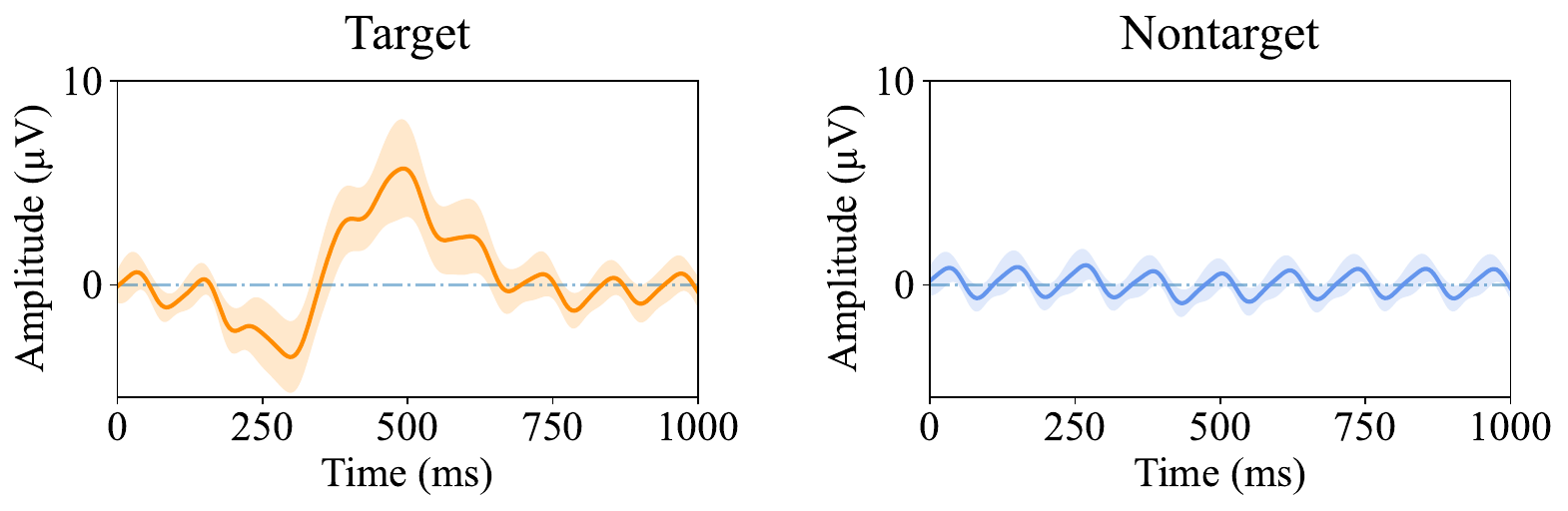}
		} 
		\subfigure[Pz waveform]{
			\includegraphics[width=0.46\linewidth]{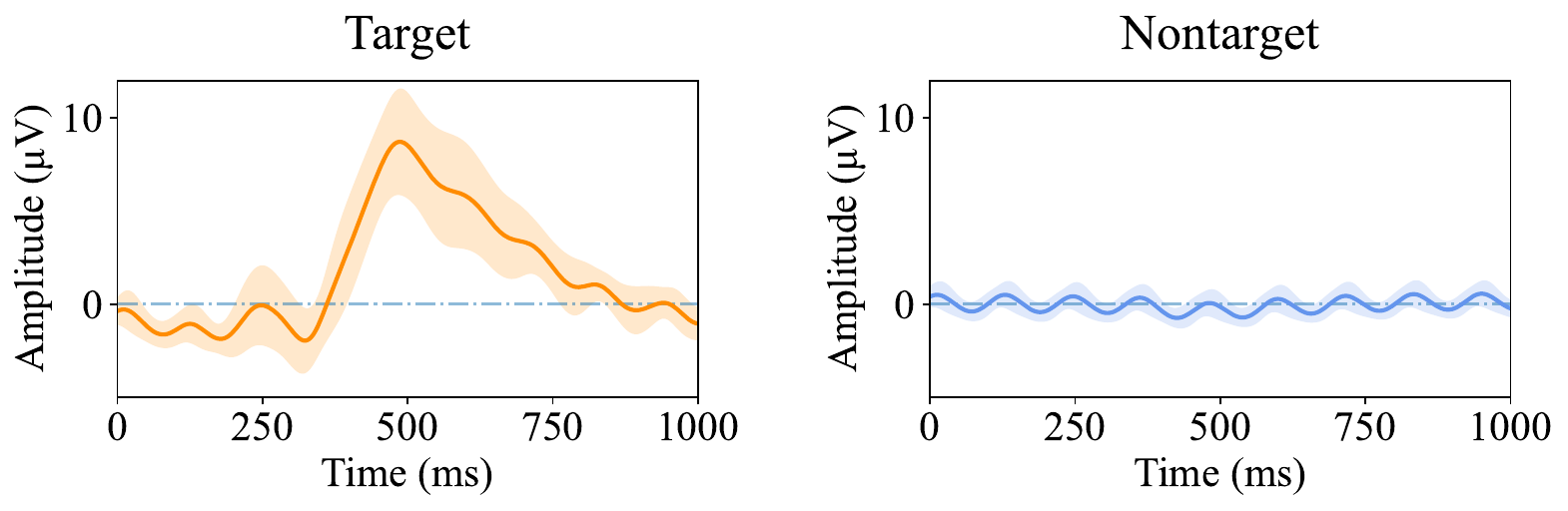}
		}
		\subfigure[Oz spectrogram image]{
			\includegraphics[width=0.47\linewidth]{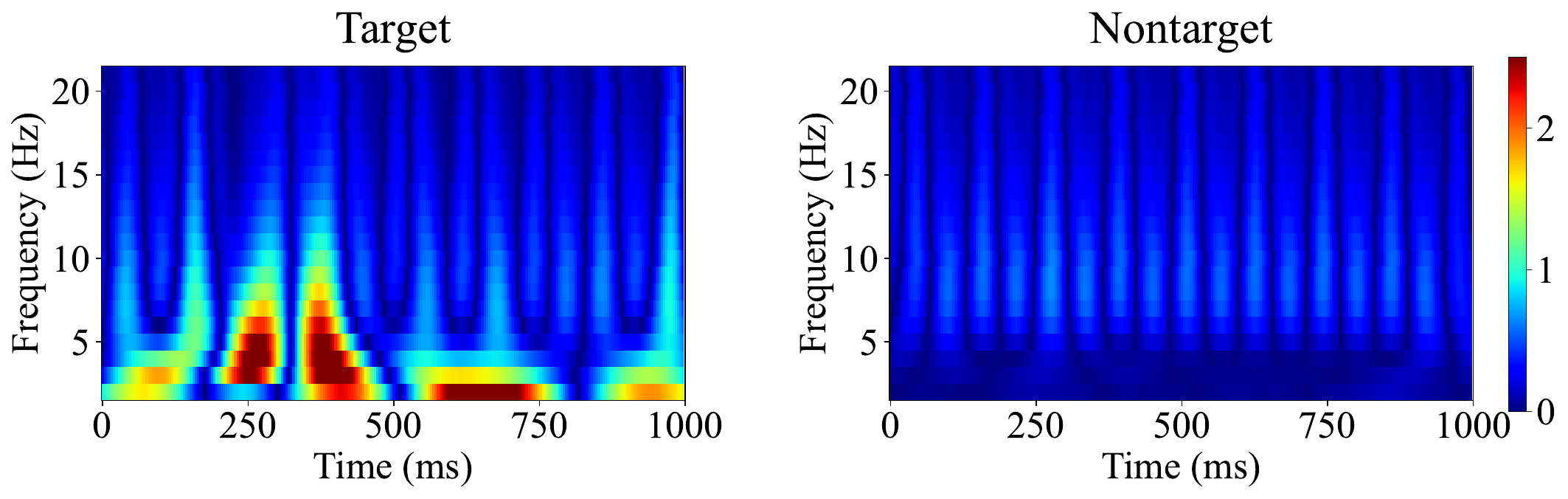}
		}
		\subfigure[Pz spectrogram image]{
			\includegraphics[width=0.47\linewidth]{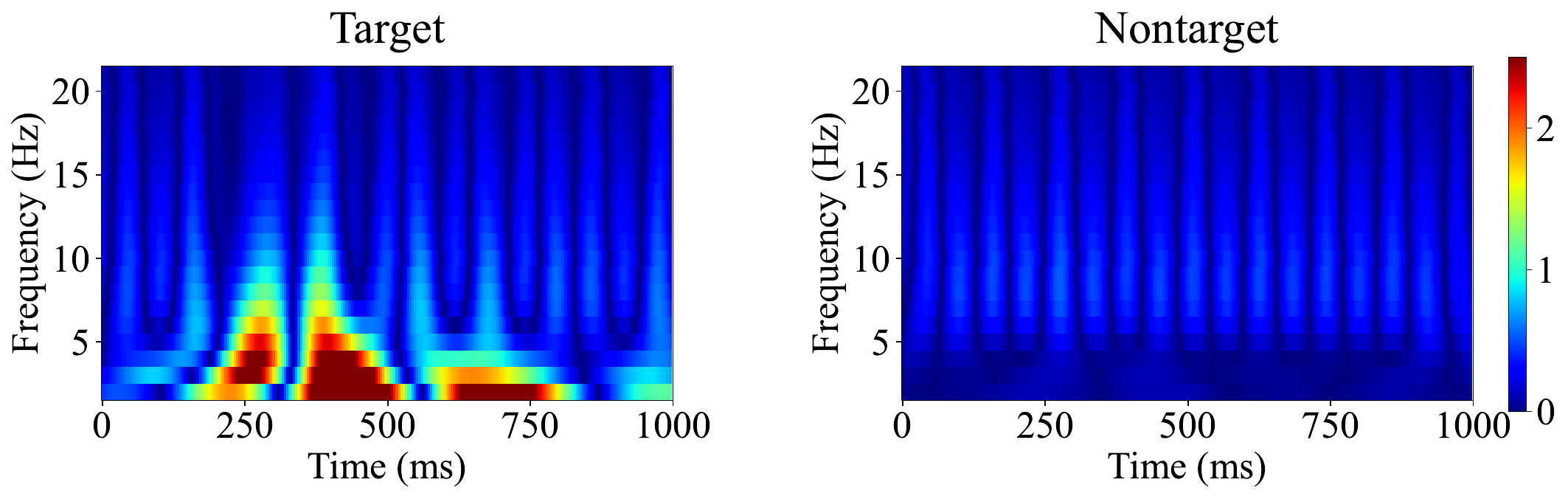}
		}
		\caption{Grand-average EEG waveforms and spectrogram images from Oz and Pz channels in Task plane. (a), (b) The target and nontarget EEG waveforms from Oz and Pz channels. (c), (d) The target and nontarget spectrogram images from the Oz and Pz channels. Shaded regions in (a) and (b) represent standard deviation.}
        \label{waveforms}
    \end{figure*}

    We display the grand-average EEG waveforms and spectrogram images of Task plane at Oz and Pz channels in Fig. \ref{waveforms}. In Fig. \ref{waveforms} (a) and (b), it is observed that the waveform of the target EEG signals exhibits the negative peak after 200 ms and the positive peak after 300 ms, corresponding to the N200 and P300 components, respectively. Nontarget EEG signals mainly present a harmonic response induced by the 10 Hz picture flicker. The target and nontarget responses of EEG signals are consistent with previous studies \cite{polich2007updating, patel2005characterization}. 
    
    As shown in Fig. \ref{waveforms} (c) and (d), the spectrogram images of the target EEG signals demonstrate high power within the 0-10 Hz frequency band at around 250 ms, as well as in the interval of 350 ms to 500 ms. Additionally, there is high power in the 0-4 Hz frequency band within the interval of 600 ms to 750 ms. Conversely, since the nontarget EEG signal is a 10 Hz harmonic response, the spectrogram images of nontarget EEG signals exhibit weakly high power which is primarily concentrated in the 5-15 Hz frequency band with 10 Hz as the center frequency at all times. These phenomena are consistent with previous studies \cite{makeig2004mining, kolev1997time}. Thus, there are discrepancies between the spectrogram images of target and nontarget EEG signals, which are primarily concentrated within the low-frequency band. These discrepancies can also be utilized for classification. 

    \subsubsection{Feature Visualization}
\begin{figure*}[!htbp]
		\centering
		\subfigure[Data]{
			\includegraphics[width=0.23\linewidth]{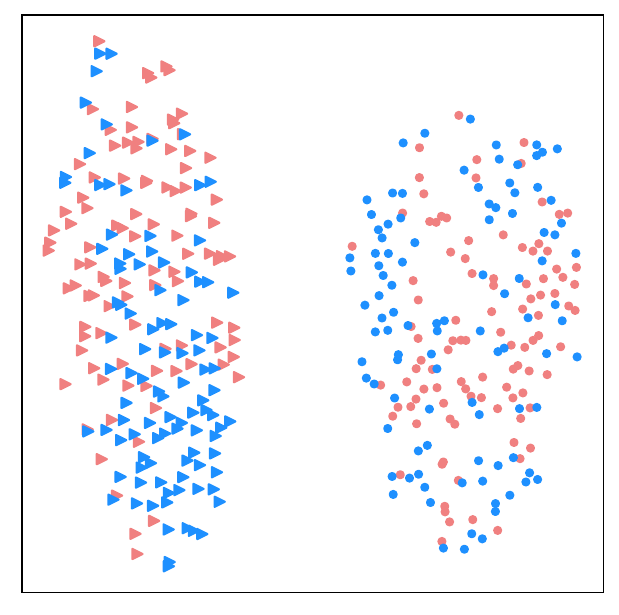}
		} 
		\subfigure[Feature extractor]{
			\includegraphics[width=0.23\linewidth]{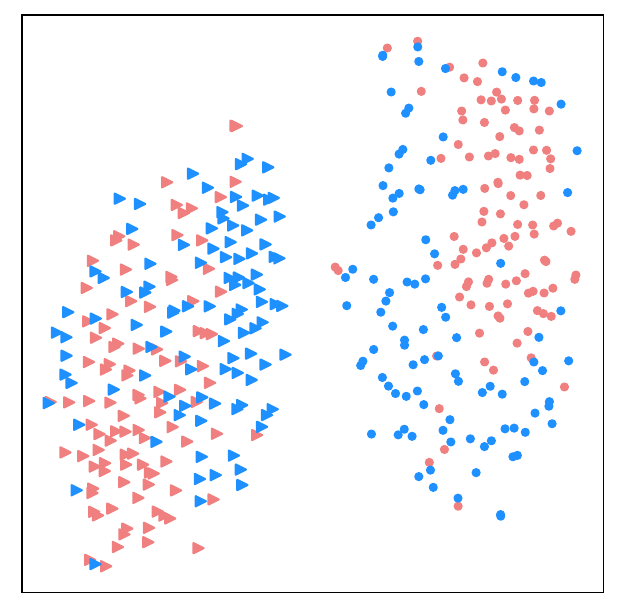}
		}
		\subfigure[Cross-view interaction module]{
			\includegraphics[width=0.23\linewidth]{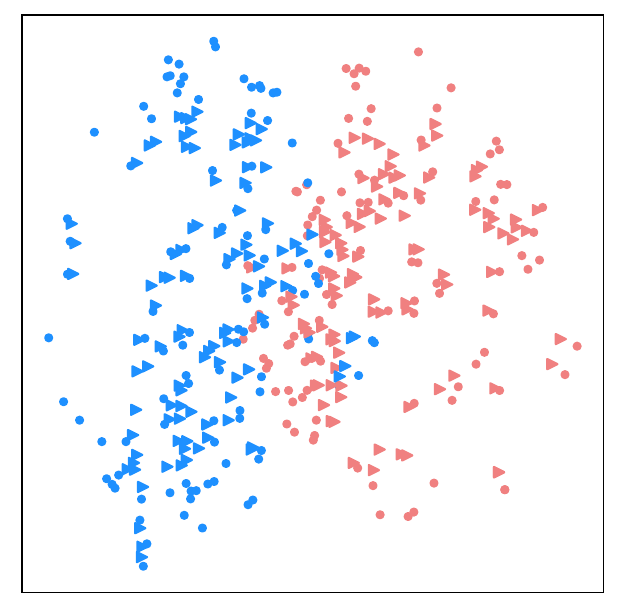}
		}
		\subfigure[Fusion module and adapter]{
			\includegraphics[width=0.23\linewidth]{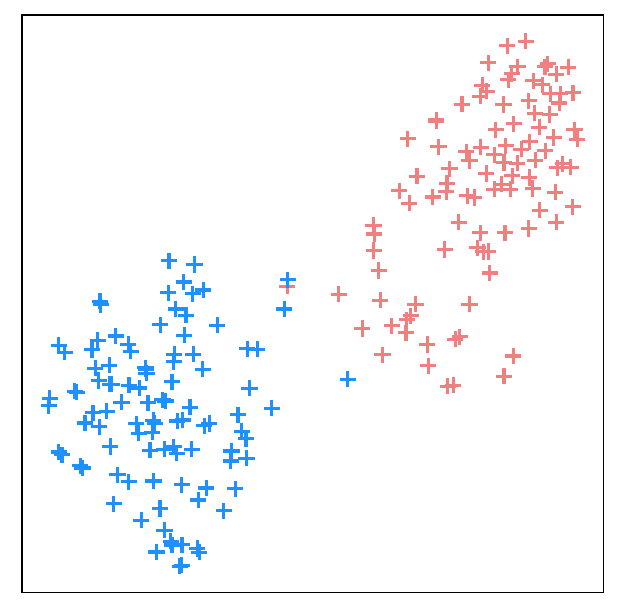}
		}
        \subfigure{
			\includegraphics[width=0.6\linewidth]{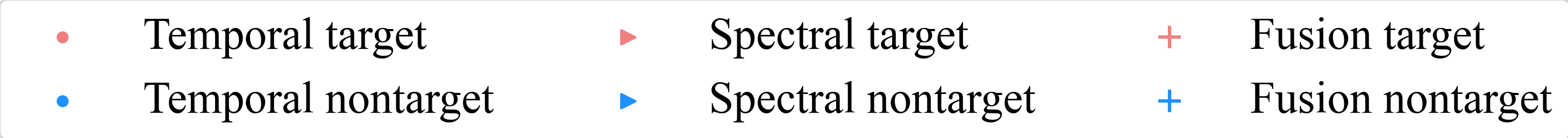}
		}
		\caption{The t-SNE visualization results of (a) raw data, (b) the output of feature extractor, (c) the output of cross-view interaction module, and (d) the output of fusion module and subject-specific adapter in TSformer-SA. The red and blue dots indicate the temporal view of target samples and nontarget samples respectively. The red and blue triangles indicate the spectral view of target samples and nontarget samples respectively. The red and blue plus signs indicate fusion features of target samples and nontarget samples respectively.}
        \label{tSNE}
    \end{figure*}

    We employ the t-distributed Stochastic Neighbor Embedding (t-SNE) \cite{van2008visualizing} to separately project the input data (see Fig. \ref{tSNE} (a)), the output features of feature extractor (see Fig. \ref{tSNE} (b)), the output features of cross-view interaction module (see Fig. \ref{tSNE} (c)), and the output features of fusion module and subject-specific adapter (see Fig. \ref{tSNE} (d)) into a two-dimensional feature space and plotted the scatter plot. We conduct this experiment on the first subject in Task Plane. The TSformer-SA is pre-trained on all other subjects in the Task plane and fine-tuned on the data from the first 4 blocks of the first subject. The remaining blocks are used for visualization. We downsample the nontarget class data in the test set to match the number of samples in the target class for better visualization. 
    
    As illustrated in Fig. \ref{tSNE} (a), the input data clusters within the temporal view and spectral view, but both temporal view and spectral view exhibit obvious overlap between target and nontarget classes. After the feature extraction by the feature extractor in each view, the overlap between target and non-target classes within each view is reduced (see Fig. \ref{tSNE} (b)). This indicates that the feature extractor can capture distinctive features from both views. In Fig. \ref{tSNE} (c), following the interaction of two-view features in the cross-view interaction module, the features between temporal and spectral views are aligned and become more closely in the feature space, while the features of target and nontarget classes tend to be linearly separated. This suggests that the cross-view interaction module can effectively extract common representations to reduce the feature gap between the two views. Subsequently, Fig. \ref{tSNE} (d) reveals the linear separation between the output features by the fusion module and the subject-specific adapter for the two classes. The fusion module and adapter successfully fuse the features from both views as more distinct fusion features and adapt to the data distribution of new subjects, which achieves superior decoding performance. Thus, our proposed model can effectively realize discriminative feature extraction and feature fusion between the two views for final classification.  
 
    \subsubsection{Subject Task-Related Feature Extraction}
\begin{figure*}[!htbp]
		\centering
		\includegraphics[width=0.9\linewidth]{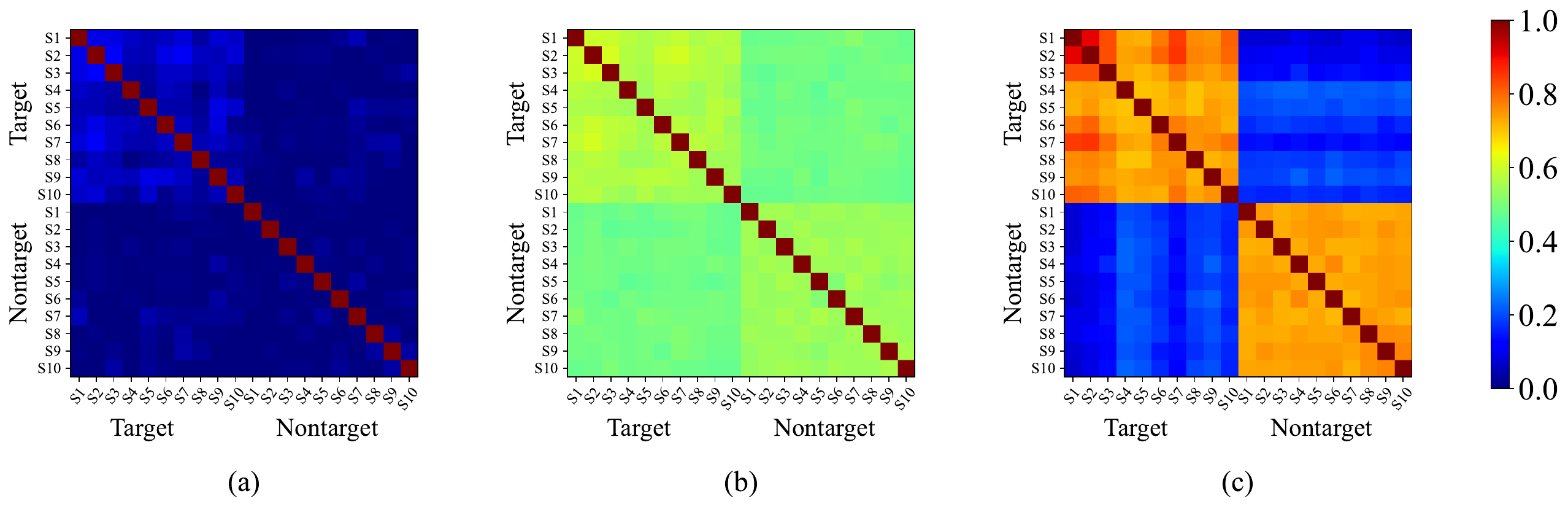}
		\caption{The cosine similarity between features extracted from target and nontarget samples from the first ten subjects in Task plane. (a) The EEG temporal signals. (b) The temporal view output from the cross-view interaction module. (c) The output from the fusion module.}
        \label{task related features}
    \end{figure*}

    This experiment is conducted to investigate the ability of our model to extract task-related representations from EEG signals across different subjects. We select the first ten subjects from Task plane. The TSformer is trained on the remaining subjects in the Task plane and directly tested on the first ten subjects. Cosine similarity is employed and visualized to assess the feature distance between the EEG temporal signals (see Fig. \ref{task related features} (a)), the temporal view output from the cross-view interaction module (see Fig. \ref{task related features} (b)), and the output from the fusion module (see Fig. \ref{task related features} (c)) of these different subjects. A target sample and a nontarget sample are randomly selected from each subject to calculate the cosine similarity. This process is repeated 1000 times and the results are averaged to draw the heatmap of similarity matrix.
    
    In Fig. \ref{task related features} (a), it can be observed that due to variations between different subjects and classes, EEG temporal signals exhibit very low similarity among subjects, with only weak similarities in the target data from different subjects. After feature extraction,  the similarity of features among subjects increases (see Fig. \ref{task related features} (b)). However, there are also notable similarities among features from different classes, which may increase FPR in classification. Figure \ref{task related features} (c) illustrates that the similarity between features from different subjects within the same class is significantly enhanced compared to Fig. \ref{task related features} (b) and is much higher than the similarity between features from different classes. This improvement strengthens the TPR while reducing FPR in classification. These results demonstrate the model's ability to effectively extract common task-related features across subjects through pre-training on data from various subjects.

    \subsection{Classification Performance with Different Number of Training Blocks}
    We conduct this experiment to investigate the performance of TSformer-SA under different amounts of training data from new subjects. In this experiment, we select HDCA, PPNN, TCN-T, and EEG-conformer as comparison methods, which achieve superior performance in subject-dependent decoding experiments among comparison methods. The models are trained with data from the first $b$ $(b=4,3,2,1)$ blocks of each subject. For Task plane and people, the last 6 blocks serve as the test set for each amount of training data, while for Task car, the last 1 block is used as the test set because each subject has only 5 blocks. Moreover, we also compare the performance of TSformer without fine-tuning on new subjects' data (i.e. subject-independent decoding) and that of the comparison methods using different amounts of training data from new subjects.

\begin{figure*}[!htbp]
		\centering
		\subfigure[Task plane]{
			\includegraphics[width=0.31\linewidth]{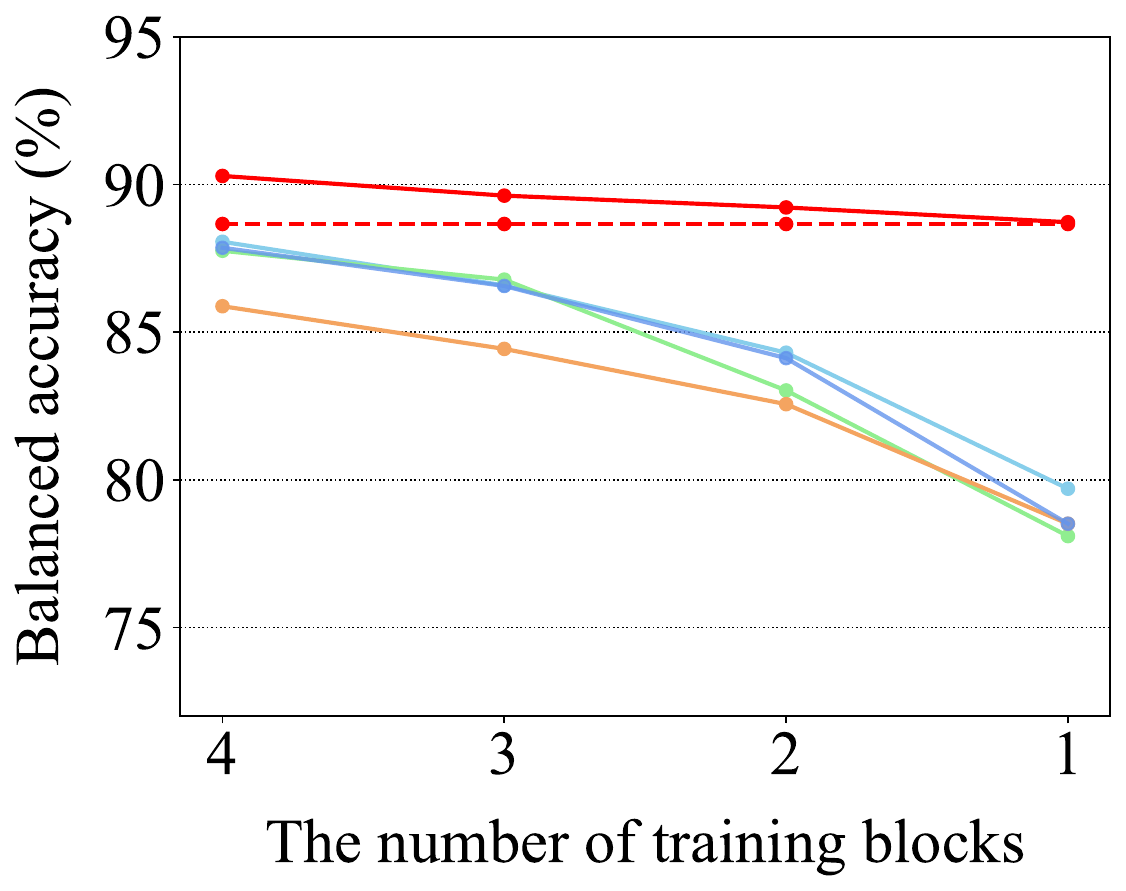}
		}
		\subfigure[Task car]{
			\includegraphics[width=0.31\linewidth]{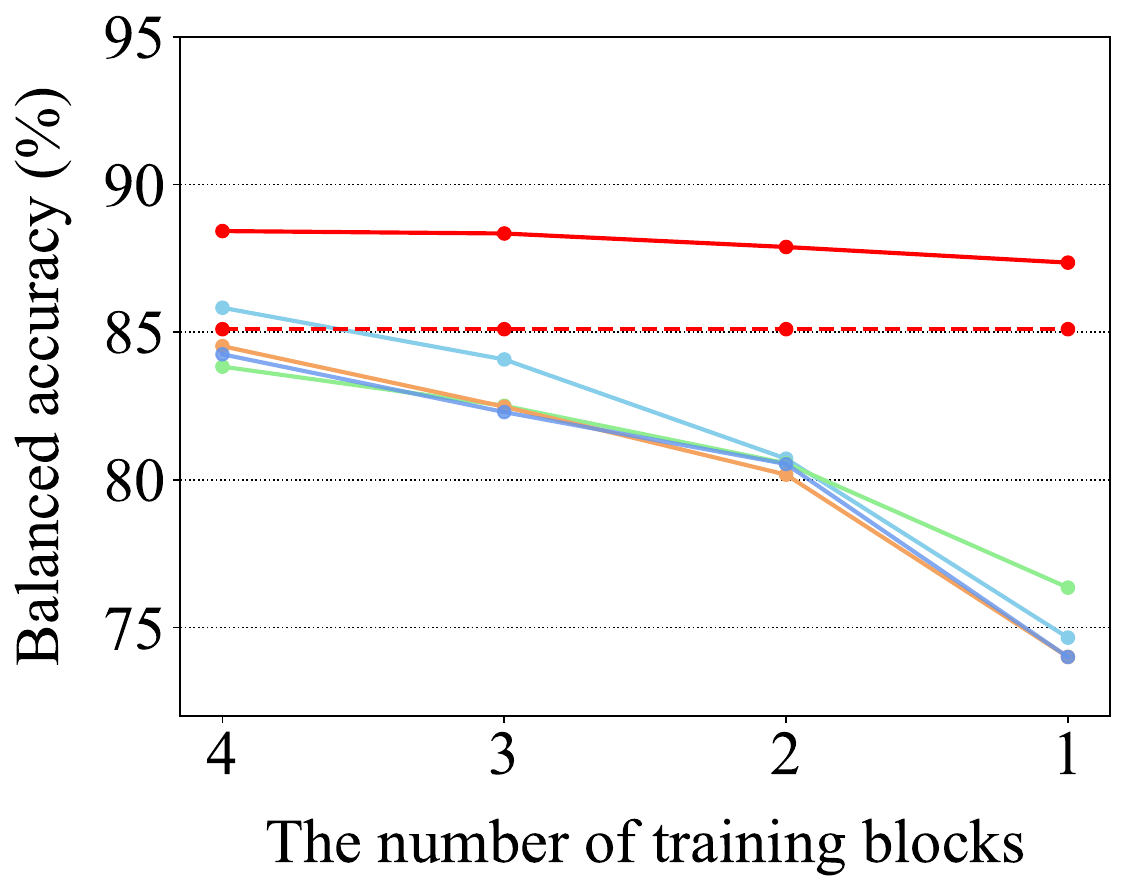}
		}
		\subfigure[Task people]{
			\includegraphics[width=0.31\linewidth]{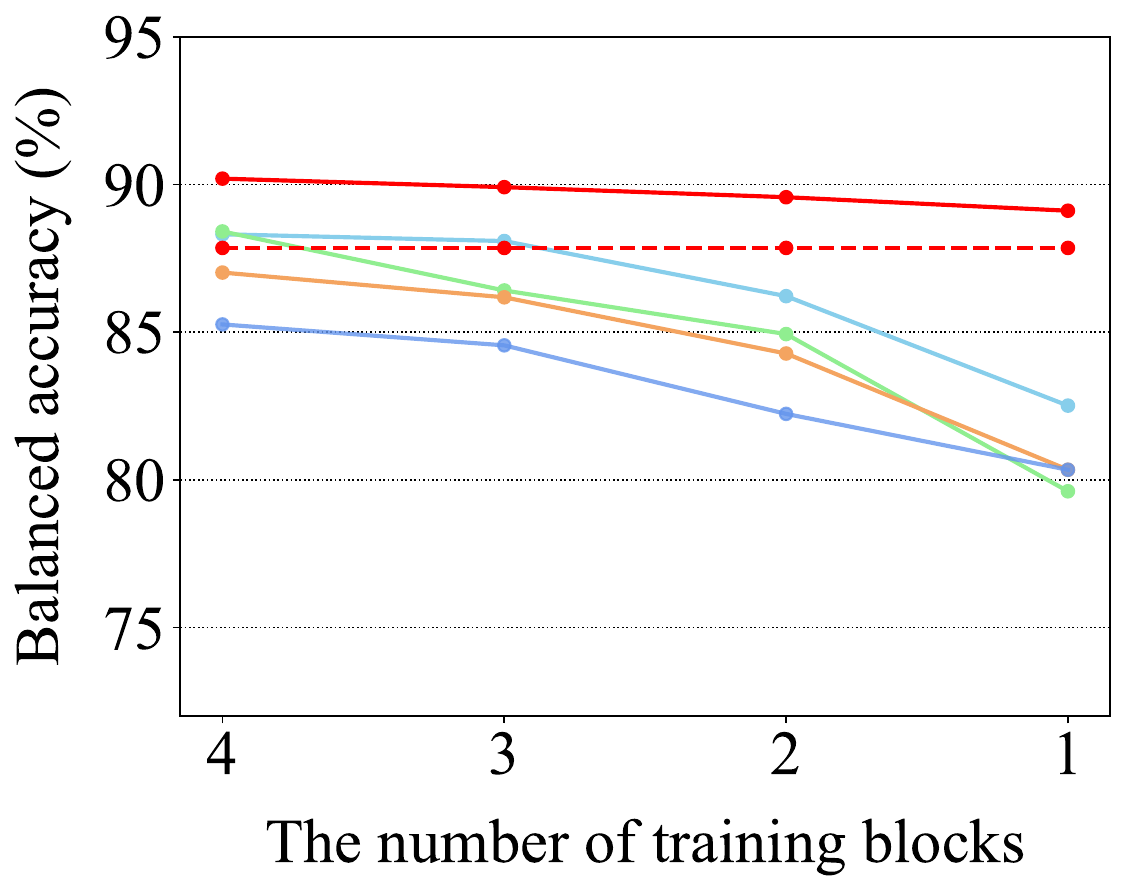}
		}
        \subfigure{
            \centering
			\includegraphics[width=0.9\linewidth]{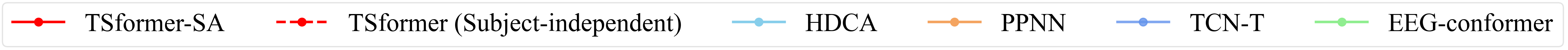}
		}\vspace{-0.2cm}
		\caption{The classification balanced accuracy of TSformer-SA and comparison methods using the various number of training blocks from the new subjects in (a) Task plane, (b) Task car, and (c) Task people.}
        \label{traing blocks}
    \end{figure*}

    Figure \ref{traing blocks} illustrates the experimental results evaluated by BA for Task plane, Task car, and Task people. In each task, the two-way repeated measures ANOVA indicate significant main effects of methods (all tasks: $p<0.001$) and the number of training blocks (all tasks: $p<0.001$), along with a significant interaction effect (all tasks: $p<0.001$). For each number of training blocks in each task, post-hoc tests reveal that TSformer-SA significantly outperforms other comparison methods (all tasks: $p<0.01$). For each comparison method in each task, when the training data is reduced by one block, the BA of the model decreases significantly (4 blocks vs. 3 blocks: $p<0.05$; 3 blocks vs. 2 blocks: $p<0.05$; 2 blocks vs. 1 block: $p<0.05$). In particular, with a reduction to 1 training block, the BA of comparison methods decreases by over 6.5\% compared to that of the models trained with 4 blocks. Conversely, the BA of TSformer-SA remains stable with each reduction of one block training data in each task, and the BA of TSformer-SA fine-tuned with one block results in a reduction of less than 1.5\% compared to that of TSformer-SA fine-tuned with four blocks. Therefore, these results indicate that TSformer-SA consistently achieves the best performance across different amounts of training data and its performance remains robust when the amount of training data is reduced. 

    Furthermore, the subject-independent decoding performance of TSformer is significantly superior to that of other comparison methods using two blocks of training data in each task (all tasks: $p<0.001$) and has no statistical difference compared to the best comparison methods (HDCA) using four blocks training data (Task plane: $p=0.57$; Task car: $p=0.94$; Task people: $p=0.64$). Thus, TSformer without fine-tuning on the new subjects' data can achieve a comparable performance to other comparison methods using four blocks of training data from new subjects. This allows TSformer-SA to skip the preparation procedure and be directly applied to RSVP decoding for new subjects. 

    \subsection{Computational Complexity in Preparation Procedure of BCI Systems}
    Before utilizing the BCI system, new subjects need to go through a preparation process which primarily consists of two parts: collecting labeled training data from new subjects and training individual models. In practical BCI applications, reducing preparation time is crucial for ensuring rapid deployment \cite{li2022tff,wei2022erp,wei2020reducing}. The results in Section 6.6 indicate that TSformer-SA requires less training data from new subjects while achieving comparable performance to the comparison methods, which can shorten the time spent on collecting new subjects' data during the preparation procedure of the RSVP-BCI system. In this part, we evaluate the time consumption of training an individual TSformer-SA during the preparation procedure.

    \setlength{\tabcolsep}{3.0mm}{
	\begin{table}[htbp]
            \footnotesize
            \renewcommand\arraystretch{1.3}
            \centering
            \caption{The number of trainable parameters and training time of TSformer-SA and compared methods in preparation procedure (mean).}
            \label{computational complexity}
            \begin{tabular}{ccc}
                \toprule[1.2pt]
                \textbf{Methods}& \textbf{Trainable Parameters} & \textbf{Training Time (s)}\\
                \midrule
                EEGNet& $\underline{9.02 \times 10^{3}}$ &$\mathbf{36.11 \pm 0.91}$  \\
                PPNN&  $17.03 \times 10^{3}$ &40.16 $\pm$ 1.09\\
                EEG-conformer&  $344.71 \times 10^{3}$ &42.62 $\pm$ 0.95\\
                TSformer-SA & $\mathbf{4.86 \times 10^{3}}$  &$\underline{37.02 \pm 0.69}$   \\
                \bottomrule[1.2pt]
            \end{tabular}
            \begin{tablenotes}
                \item \footnotesize The best results are highlighted in bold and the second-best results are denoted by underline.
            \end{tablenotes}
	\end{table}}
    
    We calculate the number of trainable parameters and training time for TSformer-SA. For comparison, we list the results of EEGNet and PPNN, which possess fewer parameters than other CNN-based comparison methods. Additionally, we include EEG-conformer, which is the Transformer-based model with the fewest parameters. We take the Task people as an example since it has the highest number of samples per block among the three tasks. For each subject, the data from the first block are utilized for fine-tuning, where 80\% of the data are randomly selected for training and the rest for validation. A training duration of 50 epochs per subject is conducted to determine the total training time, which is then averaged across subjects for the final results. The experiments are performed on an Ubuntu server with the Intel Core i7-7700 2.10GHz CPU and a NVIDIA GTX 2080 GPU.
    
    As shown in Table \ref{computational complexity}, EEGNet has the fewest trainable parameters among the comparison methods. TSformer-SA possesses the fewest trainable parameters, accounting for only 54\% of the parameter count in EEGNet. Notably, the number of trainable parameters in TSformer-SA is merely 1.5\% of that in EEG-conformer. The average training time for TSformer-SA is 37.92 seconds, which differs by less than 1 second compared to EEGNet with the least training time. Conversely, the methods in \cite{wei2020reducing} using adversarial transfer learning strategies takes 4.8 minutes of training time, which takes about eight times longer to that of TSformer-SA. Thus, our proposed two-stage training strategy ensures that only the parameters of the subject-specific adapter are updated using new subject's data during the preparation stage, which leads to a reduction in both computational costs and hardware requirements.
    
    \subsection{The Impact of Pre-training Data Sources}
    In the two-stage training strategy for TSformer-SA, both the existing subjects in the pre-training stage and the new test subjects in the fine-tuning stage come from the same RSVP task. In the same RSVP task, the target remains constant and stimulus images are obtained from the same scene. In this section, we conduct an experiment to assess the performance of TSformer-SA pre-trained and tested with data from different RSVP tasks. Specifically, each task is employed as the test task in turn, and the remaining two tasks are used separately as the pre-training tasks. Firstly, TSformer-SA is pre-trained on data from all subjects in the pre-training task. Subsequently, for each subject in the test task, we fine-tune the subject-specific adapter on their first four blocks data and test TSformer-SA on the remaining blocks data of them. The averaged BA of TSformer-SA on data from each test subject is shown in Table \ref{cross task}.

    The results on the diagonal of the table represent the BA of TSformer-SA pre-trained and tested with the same RSVP task, while the results on the remaining positions are the BA of TSformer pre-trained and tested with different RSVP tasks. In each test task, the one-way repeated ANOVA reveals a significant main effect on different pre-training tasks (all test tasks: $p<0.001$). When the test task is Task plane and car respectively, there is no statistical difference between the BA of the TSformer-SA pre-trained with Task plane and car, but they are both significantly higher than the BA of TSformer-SA pre-trained with the Task people (all: $p<0.05$). When the test task switches to Task people, the BA of the TSformer pre-trained with Task people is significantly better than that of the TSformer pre-trained with Task plane and car (all: $p<0.05$). However, the BA fluctuations of TSformer-SA pre-trained on different tasks are only 1.32\%, 1.30\%, and 1.08\% for Task plane, Task car, and Task people as test tasks, respectively. Meanwhile, TSformer-SA pre-trained on different tasks consistently outperforms the best comparison method in subject-dependent decoding on each test task (all tasks: $p<0.05$). For instance, when the Task plane serves as the test task, TSformer-SA pre-trained on the Task plane, car, and people all achieve significantly superior performance than HDCA which achieves the highest BA (88.05\%) on the Task plane in subject-dependent decoding among all comparison methods (all: $p<0.05$). 

    \setlength{\tabcolsep}{3.0mm}{
	\begin{table}[htbp]
            \footnotesize
            \renewcommand\arraystretch{1.3}
            \centering
            \caption{The balanced accuracy of TSformer-SA pre-trained and tested with data from different RSVP tasks (mean).}
            \label{cross task}
            \begin{tabular}{clll}
                \toprule[1.2pt]
                \multirow{2}{*}{\textbf{Pre-training task}}& \multicolumn{3}{c}{\textbf{ Test task}}\\
                \cmidrule{2-4}
                & \textbf{Task plane} & \textbf{Task car} & \textbf{ Task people}\\
                \hline
                \textbf{Task plane} &  $\underline{90.29}$& $\mathbf{88.94}$ &  $89.12^{\star\star\star}$  \\
                \textbf{Task car} & $\mathbf{90.41}$ & $\underline{88.42}$ & $\underline{89.39}^{\star}$ \\
                \textbf{Task people} & $89.09^{\star\star}$ &   $87.64^{\star\star\star}$& $\mathbf{90.20}$   \\
                \bottomrule[1.2pt] 
            \end{tabular}
            \begin{tablenotes}
                \item \footnotesize The asterisks in the table indicate a significant difference between this result and the best result by paired t-tests ($^{\star}$ $p<0.05$, $^{\star\star}$ $p<0.01$, $^{\star\star\star}$ $p<0.001$). The best results are highlighted in bold and the second-best results are denoted by underline.
            \end{tablenotes}
    \end{table}}

    These results demonstrate that although the performance of the model can be affected when the pre-training task differs from the test task, TSformer-SA consistently achieves stable and excellent performance. Consequently, in practical applications, the pre-training stage of TSformer-SA will not be limited by insufficient existing data from the same RSVP task performed by new subjects. Instead, it can leverage available data from other RSVP tasks for pre-training to realize efficient decoding on new test subjects.
 
    \subsection{Sensitive Analysis of Wavelet Basis Function}
\begin{figure}[!htbp]
		\centering
		\includegraphics[width=1.0\linewidth]{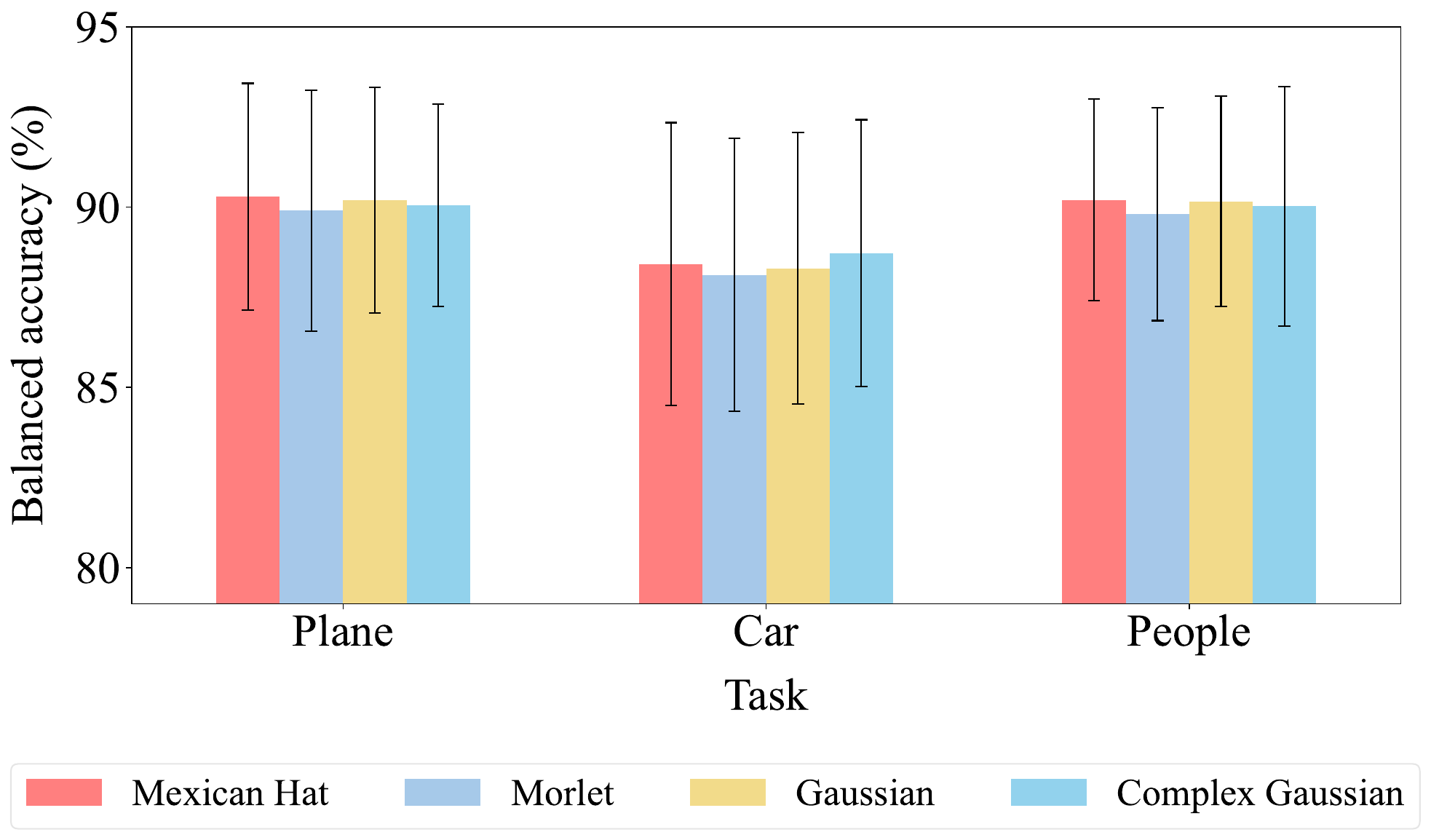}
		\caption{The classification balanced accuracy of TSformer-SA using spectrogram images obtained by CWT with different wavelet basis functions.}
        \label{sensitive}
    \end{figure}

    The spectral view input of TSformer-SA is obtained through CWT, which relies on two important factors: the wavelet basis function and the scale. Considering that the EEG characteristics in RSVP decoding are primarily concentrated in the low-frequency band, we apply a band-pass filter of 0.5-15 Hz during the data preprocessing stage. Thus, we appropriately expand the frequency band range and set the scale to 20, which can capture EEG signal characteristics within the interval between 0 Hz and 20 Hz. To explore the influence of different wavelet basis functions on decoding performance, we conduct a sensitivity analysis. Specifically, we employ representative wavelet basis functions, such as Mexican hat, Morlet, Gaussian, and Complex Gaussian, to perform CWT on the EEG signals as the spectral view input of TSformer-SA for RSVP decoding. The results are illustrated in a bar chart depicted in Fig. \ref{sensitive}. The one-way repeated ANOVA in each task reveals no main effects of different wavelet basis functions (Task plane: $p=0.52$, Task car: $p=0.71$, Task people: $p=0.40$). This indicates that the performance of our model is robust to the different wavelet basis functions. 
    
    \subsection{Limitation and Future Work}
    Although our method achieves superior performance using EEG signals, it does not take advantage of multi-modal features such as eye movement signals and stimulus image features, which may further enhance RSVP decoding performance. In future research, we can explore a multi-modal learning approach that integrates EEG signals, eye movement data, and stimulus image characteristics to further enhance the decoding performance of RSVP tasks. Secondly, due to the limited dataset size, our model size is relatively small, which may restrict the decoding performance of our model across different subjects and datasets. In the future, we will try to expand the size of the dataset and then construct a pre-trained large model for RSVP tasks, which can improve the generalization ability of the decoding model across different subjects and datasets.
    
    \section{Conclusion}
    This study introduces the Temporal-Spectral fusion transformer with Subject-specific Adapter (TSformer-SA) to integrate temporal and spectral views of EEG signals for enhancing RSVP decoding performance. The TSformer-SA employs a feature extractor to capture view-specific features from EEG temporal signals and spectrogram images obtained by CWT. A cross-view interaction module is then utilized to facilitate information transfer and common representation extraction across the two views. The features from the two views are further fused in the fusion module to obtain more discriminative features for classification. The TSformer-SA is initially pre-trained on data from existing subjects to extract task-related features across different subjects and EEG views. Subsequently, the subject-specific adapter is fine-tuned to rapidly transfer the learned patterns to the decoding task on the new test subject. Experimental results demonstrate that our model significantly outperforms all the comparison methods. Moreover, our model exhibits excellent performance with limited amounts of training data from new subjects or even without training data from new subjects. It can be trained and deployed rapidly during the preparation procedure. These results indicate that our model can enhance RSVP decoding performance while minimizing the preparation time required before employing BCI systems, which further promotes BCI systems from research to practical use.

\bibliographystyle{ieeetr}
\bibliography{sample-base}

\end{document}